\documentclass{aa}

\usepackage{graphicx}   
\usepackage{txfonts}
\usepackage{subcaption}
\usepackage{natbib}

\newcommand{\sref}[1]{Section~\ref{#1}}
\newcommand{\fref}[1]{\autoref{#1}}
\newcommand{\tref}[1]{\autoref{#1}}
\newcommand{\eref}[1]{Eq.~\eqref{#1}}

\usepackage[colorlinks=true,linkcolor=blue,citecolor=blue, urlcolor=blue]{hyperref}
\makeatletter
\renewcommand*\aa@pageof{, page \thepage{} of \pageref*{LastPage}}
\makeatother

\begin{document} 

   \title{
   Swarming in stellar streams: Unveiling the structure of the Jhelum stream with ant colony-inspired computation\\
   }
   \titlerunning{Sub-structure within the Jhelum Stream proper motion space}
\author{
P.~Awad\inst{\ref{kapteyn}, \ref{bernoulli}}
\and
M. Canducci\inst{\ref{birmingham}}
\and
E. Balbinot\inst{\ref{leiden}, \ref{kapteyn}}
\and
A. Viswanathan\inst{\ref{kapteyn}}
\and
H. C. Woudenberg\inst{\ref{kapteyn}}
\and
O. Koop\inst{\ref{kapteyn}}
\and
R. Peletier\inst{\ref{kapteyn}}
\and
P. Ti{\v{n}}o\inst{\ref{birmingham}}
\and 
E. Starkenburg\inst{\ref{kapteyn}}
\and
R. Smith\inst{\ref{santiago}}
\and 
K. Bunte\inst{\ref{bernoulli}}
}

\institute{
Kapteyn Astronomical Institute, University of Groningen, PO Box 800, 9700 AV Groningen, The Netherlands
\label{kapteyn}
\newline
\email{p.awad@rug.nl}
\and
Bernoulli Institute for Mathematics, Computer Science and Artificial Intelligence, University of Groningen, 9700AK Groningen, The Netherlands
\label{bernoulli}
\and 
University of Birmingham, School of Computer Science, B15 1TT, Birmingham, United Kingdom
\label{birmingham}
\and
Leiden Observatory, Leiden University, P.O. Box 9513, NL-2300 RA Leiden, The Netherlands
\label{leiden}
\and
Universidad Technica Frederico de Santa Maria, Avenida Vicuña Mackenna 3939, San Joaquín, Santiago, Chile
\label{santiago}
}

   \date{Received \today; accepted }

 
\abstract{
The halo of the Milky Way galaxy hosts multiple dynamically coherent substructures known as stellar streams that are remnants of tidally disrupted orbiting systems such as globular clusters (GCs) and dwarf galaxies (DGs). 
A particular case is that of the Jhelum stream, which is known for its unusual and complex morphology. 
Using the available data from the \emph{Gaia} DR3 catalog, we extracted a region on the sky that contains Jhelum, and fine-tuned this selection by enforcing limits on the magnitude and proper motion of the selected stars. 
We then applied the novel Locally Aligned Ant Technique (LAAT) on the position and proper motion space of stars belonging to the selected region to highlight the stars that are closely aligned with a local manifold in the data and the stars belonging to regions of high local density. 
We find that the overdensity representing the stream in proper motion space is composed of two components, and show the correspondence of these two signals to the previously reported narrow and broad spatial components of Jhelum.
We then made use of the radial velocity measurements provided by the $S^5$ survey and confirm, for the first time, a separation between the stars belonging to the two components in radial velocity. 
We show that the narrow and broad components have velocity dispersions of $4.84^{+1.23}_{-0.79}$~km/s and $19.49^{+2.19}_{-1.84}$~km/s, and metallicity dispersions of $0.15^{+0.18}_{-0.10}$ and $0.34^{+0.13}_{-0.09}$, respectively. 
These measurements, as well as the given difference in component widths, could be explained with a probable scenario where
Jhelum is the remnant of a GC embedded within a DG where both were accreted onto the Milky Way during their infall. Although the properties of Jhelum could be explained with this merger scenario, other progenitors of the narrow component remain possible such as a nuclear star cluster or a DG. To rule these possibilities out, we would need more observational data of member stars of the stream. Our analysis shows that the internal structure of streams holds great information on their past formation history, and therefore provides further insight into the merger history of the Milky Way.}




   \maketitle
%

\section{Introduction}
\label{sec:introduction}

The history of the formation of the Milky Way's stellar halo is, in large part, caused by mergers with globular clusters (GCs) and nearby dwarf galaxies (DGs). The tidal disruption induced by the host galaxy on the orbiting systems causes a loss of mass in the form of elongated distributions of stars, also known as stellar streams \citep{HelmiEtal1999, CombesEtal1999, EyreEtal2009}. 
These stellar remnants of past mergers are excellent probes of the acceleration field of the Galaxy over the spatial range of their structure \citep{IbataEtal2002, JohnstonEtal2002, Carlberg2012}. 
The studied acceleration field can then provide constraints on the gravitational force and the distribution of dark matter within the halo. 
This in turn allows for the assessment of the standard Lambda cold dark matter ($\Lambda$CDM) cosmology by comparing the observationally informed constraints with predictions made by the assumed cosmology. 
Additionally, the elongated thin nature of these streams makes them sensitive to perturbations induced by small-scale gravitational encounters. 
Therefore, stellar streams can also be useful tools for studying dark matter sub-halos by studying the gaps in stellar streams that could be the result of an impact with these sub-halos \citep{Bovy2016, BanikEtal2018, BonacaEtal2019b, MontanariEtal2022}. 
Consequently, the search for stellar streams has grown over the years and has been successful in finding several such streams in orbit around the Galaxy \citep{OdenkirchenEtal2001, NewbergEtal2002, GrillmairEtal2006, BelokurovEtal2007, IbataEtal2018, IbataEtal2021}. 
The proliferation of these studies has also been greatly aided by the availability of large photometric surveys such as the Sloan Digital Sky Survey \citep[SDSS]{SDSS}, the Dark Energy Survey \citep[DES]{DESDR1}, and most notably the data releases from the \emph{Gaia} mission \citep{GaiaDR1, GaiaDR2, GaiaDR3}.

Of particular interest when studying stellar streams is attempting to determine what the properties of the progenitor were. 
Such information provides great insight into the masses, chemical compositions, and stellar populations of these progenitors which had been embedded in the Galaxy and played a role in its formation. 
Works such as \citet{BonacaEtal2021} and \citet{MalhanEtal2022} have attempted to group streams, GCs, and satellite galaxies in action space and trace them back to past mergers with the Milky Way. In addition to these studies based on large populations, every stream has its own past with regards to sub-halo interactions and thus merger histories. 
In this work, we focus on the stellar stream Jhelum, which was first discovered by \citet{ShippEtal2018} using data from DES and subsequently confirmed by \citet{MalhanEtal2018} using \emph{Gaia}'s second data release (DR2). The proper motion of Jhelum was further analyzed in \citet{ShippEtal2019}. The most recent studies of Jhelum have classified it to be more likely the remnant of a DG \citep{JiEtal2020, BonacaEtal2021, LiEtal2022}. Such a classification is not surprising given the wide morphology for which Jhelum is known. In \citet{BonacaEtal2019}, evidence for two spatial components of Jhelum was found. Aided by photometric measurements from DES and proper motion measurements from \emph{Gaia} DR2, \citet{BonacaEtal2019} showed that Jhelum is composed of two parallel components: a narrow dense component and a broad diffuse component beneath it. Dynamic environments induced by interactions with sub-halos, the Large Magellanic Cloud (LMC), or with an asymmetric potential of the Milky Way, for example, can also disperse the stars that were originally part of a thin stream and form wider structures as a result \citep{BonacaEtal2014, NganEtal2016, PearsonEtal2017}. \citet{WoudenbergEtal2023} have therefore suggested that Jhelum could be a stream perturbed by the orbit of the Sagittarius DG. On the other hand, other explanations are also popular. For example, a GC within a satellite galaxy that has fallen into the Galaxy's potential can create a dynamically cold stream accompanied by a wider component with a lower surface brightness \citep{Carlberg2020, Qian2022}.  

In addition to recognizing them spatially, \citet{BonacaEtal2019} found no distinguishable difference between the two components in proper motion, while \citet{ShippEtal2019} independently measured two distinct proper motion components with consistent spatial distributions. 
In this work, we build on the latter two studies and confirm the substructure within the proper motion space of Jhelum with measurements from \emph{Gaia} DR3. 
Specifically, using a novel machine learning tool, the Locally Aligned Ant Technique \citep[LAAT]{LAAT}, we distinguished two components in proper motion space that we attribute to the narrow and broad spatial components of the stream. 
We also found, for the first time, a separation between the two components in the third velocity component, namely using radial velocity measurements from the Southern Stellar Stream Spectroscopic Survey \citep[$S^5$]{LiEtal2019, LiEtal2022}. 
With this new knowledge, we provide our estimates of several properties of Jhelum such as the best-fit orbit, the velocity and metallicity dispersions, as well as the width of either component. Additionally, we attempted to constrain the more probable progenitors of the narrow and broad components of Jhelum.

This paper is organized as follows. 
\sref{sec:data} presents our preliminary data selection to locate the Jhelum stream. \sref{sec:method} describes the procedure followed using our novel methodology to isolate the stars belonging to the stream from the surrounding field stars. 
In \sref{sec:results}, we explain how we fine-tuned our selection of stars and related the two overdensities in the proper motion space that were first noticed in \citet{ShippEtal2019} to the spatial narrow and broad components of Jhelum. In \sref{sec:properties} we provide our estimates of the properties of the two components. \sref{sec:discussion} contains our discussion of the star selection criteria and the most likely merger scenario. In \sref{sec:conclusion}, we summarize our findings and suggest possible future developments.

\section{Data}
\label{sec:data}

For the selection of the area on the sky containing Jhelum, we rely on the coordinate system defined in \citet{BonacaEtal2019} and implemented in {\ttfamily gala} \citep{PriceWhelanEtal2017} to transform any set of coordinates into the Jhelum frame $(\phi_1, \phi_2)$, where $\phi_1$ is the coordinate aligned with the stream track and $\phi_2$ is the coordinate perpendicular to it. We query the \emph{Gaia} DR3 catalog \citep{GaiaDR3} within the rectangular region defined by $-5^{\circ} < \phi_1 < 30^{\circ}$ and  $-5^{\circ} < \phi_2 < 5^{\circ}$, and select all stars with parallaxes larger than 1~mas. Using {\ttfamily gala}, we then correct the proper motions of the selected stars for the solar reflex motion relying on the procedure of \citet{PriceWhelanBonaca2018} assuming a constant distance to the stream of 13~kpc. To narrow down our selection of stars belonging to the Jhelum stream, we follow \citet{BonacaEtal2019} and retain the stars belonging to the region in proper motion space defined by  $-8 < \mu_{\phi_1} / \rm mas \; yr^{-1}< -4$ and $-2 < \mu_{\phi_2} / \rm mas \; yr^{-1}< -2$, where $\mu_{\phi_1}$ and $\mu_{\phi_2}$ are the proper motions projected along $\phi_1$ and $\phi_2$ respectively.

\begin{figure}
\centering
\includegraphics[width=\columnwidth]{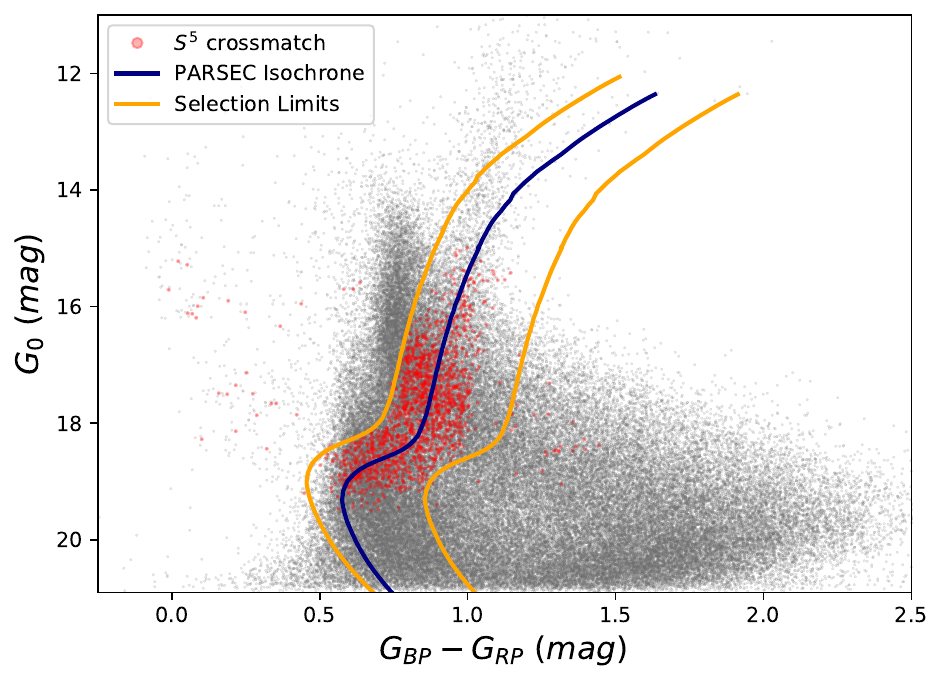}
\caption{The initial color and magnitude selection. In gray we plotted the color-magnitude diagram (CMD) of the stars that follow proper motion selection and extinction correction. Red points denote common stars between the selection in gray and the stars observed by the $S^5$ survey. Using these stars and a 12~Gyr, [Fe/H]$=-1.7$ PARSEC isochrone at 13~kpc (dark blue), we define the primary CMD selection as all stars lying in between the orange lines. The selection limit line on the right is positioned there to capture as many main sequence stars as possible while containing most stars in common with the $S^5$ survey.}
\label{fig:PolygonSelect}
\end{figure}

\begin{table*}[ht!]
\caption{Input parameters for LAAT and run information.}             
\label{table:1}      
\centering          
\begin{tabular}{c c l c c c c c}     
\hline 
Param. & Value & Definition & Run & $r$ (deg) & $N_i$ & $N_f$ & Time (min)\\ 
\hline                        
    $N_{\rm agents}$ & 500 & Number of agents \\ 
   $N_{\rm iter}$ & 10 & Number of iterations & 1 & 0.5 &37418 & 10880 & $\sim15$\\  
   $N_s$ & 2500 & Number of steps taken by each agent\\
   $r$ & 0.5$-$1 & Size of a neighborhoods radius & 2 & 0.75 & 8452 & 3414 & $\sim10$\\  
   $\kappa$ & 0.5 & Contribution of alignment vs pheromone \\
   $\beta$ & 10 & Inverse temperature for jump-probabilities & 3 & 0.75 & 2442 & 501 & $\sim10$\\  
   $\gamma$ & 0.05 & Deposited pheromone per visit per agent \\
   $\zeta$ & 0.1 & Evaporation rate per iteration \\
\hline\hline               
\end{tabular}
\tablefoot{From left to right, we give: the parameters used for running the Locally Aligned Ant Technique (LAAT), the values used for each parameter, and its definition. Note that we give the advisable range for the neighborhood radius parameter $r$ for this specific work. In the second part of the table we give: the index of the three runs, the value of $r$ used for each run, the number of stars in the input dataset, $N_i$, the number of stars remaining after the filtration based on the pheromone, $N_f$, and the time needed for each LAAT run using the unparallelized MATLAB implementation of LAAT on a machine with a 1.8~GHz $\times 8$ processor and 15.3 GiB of RAM memory.}
\end{table*}

Finally, we apply a selection in color-magnitude space to further constrain the stars more likely to belong to the stream. First, all magnitudes of the stars are corrected for extinction using the \citet{SchlegelEtal1998} dust maps and assuming a \citet{CardelliEtal1989} extinction law with $R_v$ = 3.1. The color-magnitude diagram (CMD) of the stars selected so far and corrected for extinction and reddening is shown in gray in \fref{fig:PolygonSelect}. To define a selection of stars within the CMD, we first cross-match all remaining stars with the entire Southern Stellar Stream Spectroscopic Survey $(S^5)$ catalog \citep{LiEtal2019, LiEtal2022} to locate the stars which have radial velocities measured by the survey. The latter is a spectroscopic survey that makes use of the 3.9~m Anglo-Australian Telescope's positioner and AAOmega spectrograph combined with the photometry of the Dark Energy Survey DR1 \citep[DES]{DESDR1} and precise proper motions from \emph{Gaia} DR2 \citep{GaiaDR2} to map, in detail, the properties of the stellar streams in the southern hemisphere of the Galaxy. The survey specifically targets stars in the Galactic halo with a focus on stellar streams including Jhelum. Thus, mapping the position of these stars in the CMD allows us to better constrain our color and magnitude selection. The stars in common between our selection and the $S^5$ catalog are shown in red in \fref{fig:PolygonSelect}. 
We also define a 12 Gyr, [Fe/H]$=-1.7$ PARSEC isochrone at a distance of 13~kpc which fits the CMD reasonably well following \citet{WoudenbergEtal2023}. This isochrone is therefore shown in \fref{fig:PolygonSelect} in dark blue as a primary reference for the location of the red giant branch (RGB), sub-giant, turn off and main sequence (MS) stars belonging to the stream. Using the CMD of the stars cross-matched with the $S^5$ catalog and the location of this isochrone, we define a region in the CMD bordered by the orange lines, which contains all stars matching with the $S^5$ survey and following the trend of the mentioned isochrone. We keep all stars within the defined region and use them as the starting selection for our subsequent analysis. The purpose of choosing such a wide region, as opposed to works such as \citet{BonacaEtal2019}, \citet{SheffieldEtal2021}, \citet{WoudenbergEtal2023}, or \citet{ViswanathanEtal2023} is to be as inclusive as possible of the stars belonging to Jhelum, and from there begin fine-tuning this selection to detect stars that belong to the stream with high probability. The narrowing-down procedure is explained in \sref{sec:method} and further discussed in \sref{sec:discussion}.

\section{Method for stream extraction}
\label{sec:method}

We now begin the procedure of fine-tuning the color-magnitude selection to create a high-purity sample of stars belonging to the Jhelum stream. With that purpose, we employ the Locally Aligned Ant Technique (LAAT) first introduced in \citet{LAAT}, and then grouped into a toolbox of manifold (structure) extraction and modeling algorithms in \citet{CanducciEtal2022} and \citet{AwadEtal2022}. The main purpose of LAAT is to highlight the contrast between low and high-density regions in a given point cloud, 
as well as the detection of regions that are closely aligned with a defined structure (low-dimensional smooth manifolds) within the spatial distribution of data points (such as 
stars, gas particles, and simulated dark matter particles). 

The algorithm operates based on the idea of Ant Colony Optimization \citep[ACO]{ACObook} whereby a number of agents or "ants" are distributed in the data (e.g., in the position space of a given point cloud dataset) and a random walk is initiated in that space. A defined quantity termed the ``pheromone" is artificially deposited on the data points visited by the agents\footnote{Note that in the original ACO formulation pheromone would be deposited on path segments joining pairs of data points. However, for our purposes, depositing pheromone on individual data points is sufficient and computationally much more efficient.} and is associated with an ``evaporation rate." The latter reduces the quantity of pheromone on the data points over time. Ants, in their random walk, prefer paths with more accumulated pheromone thus implementing a form of a "positive feedback loop." Points visited more frequently will accumulate more pheromone that would take longer to evaporate, thus attracting even more ants. Such a positive reinforcement mechanism distinguishes the ant colony from a random walk defined by a Markov chain without the pheromone mechanism. There, the visitation frequency of the points would be given by the stationary distribution of the Markov chain. For a more formal treatment see \citet{Mohammadi2022}.
One can shape the random walk to concentrate on different forms of spatial structures. In our case, we would like to emphasize the points aligned with low-dimensional manifolds in the data cloud. Hence, during the walk, and given a data point on which an agent is currently located, principal component analysis (PCA) is performed within the neighborhood with a chosen radius $r$ and centered at the location of the point, to distinguish the main directions along which the data points are distributed
in that neighborhood. 
The agent is then incentivized to jump toward points within the neighborhood that 1) align with the dominant direction of distribution of the data points, and 2) have accumulated larger amounts of pheromone as the walk is allowed to continue. 
Since data points belonging to a structure in the data show more directional alignment and have a larger local density than a random distribution of points, the agents have a higher probability of visiting and depositing pheromone on structures embedded within the data. 
As the walk is allowed to progress over multiple iterations, the pheromone will accumulate on data points belonging to the embedded structures, and evaporate in sparser, less directionally aligned regions. 
Finally, a threshold can be enforced on the pheromone quantity to extract the detected structures from the data. For a detailed mathematical description of the algorithm, see \citet{LAAT}.
 \begin{figure*}[ht!]
     \centering
     \begin{subfigure}[t]{\textwidth}
     \centering
     \includegraphics[width=0.95\textwidth]{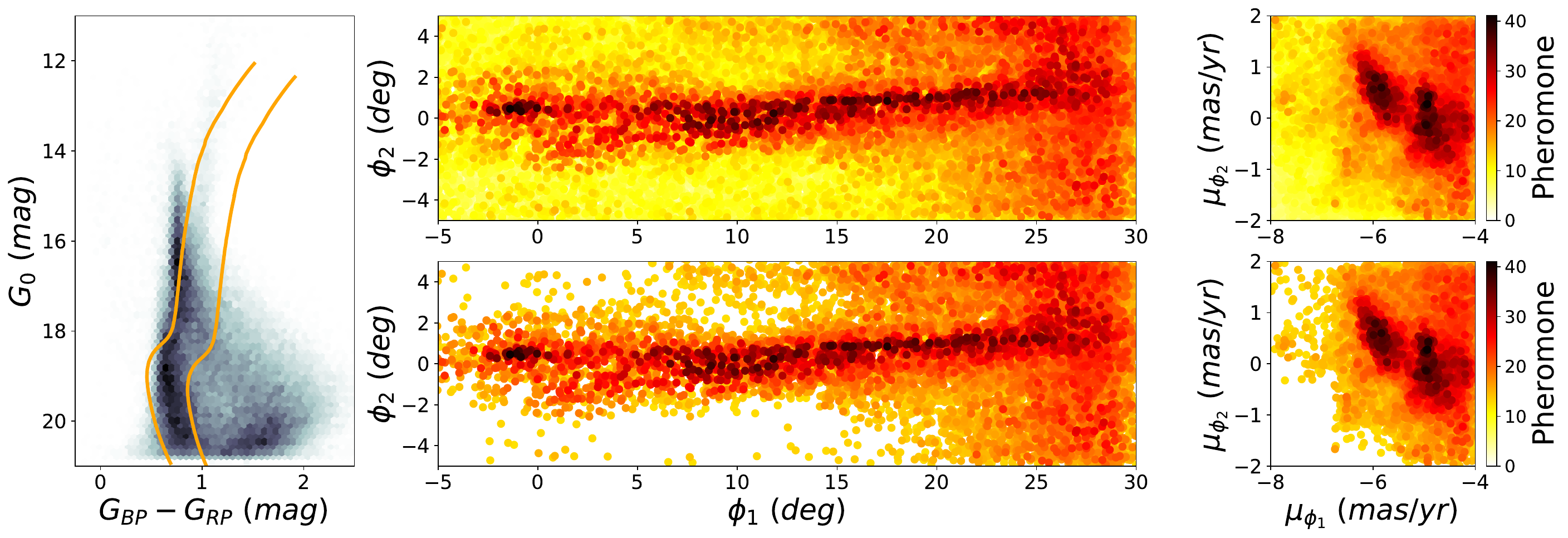}
     \caption{First run of LAAT applied with a threshold of $ 30\%$.}
     \label{fig:LAAT1}
     \end{subfigure}
     \begin{subfigure}[t]{0.95\textwidth}
     \centering
     \includegraphics[width=\textwidth]{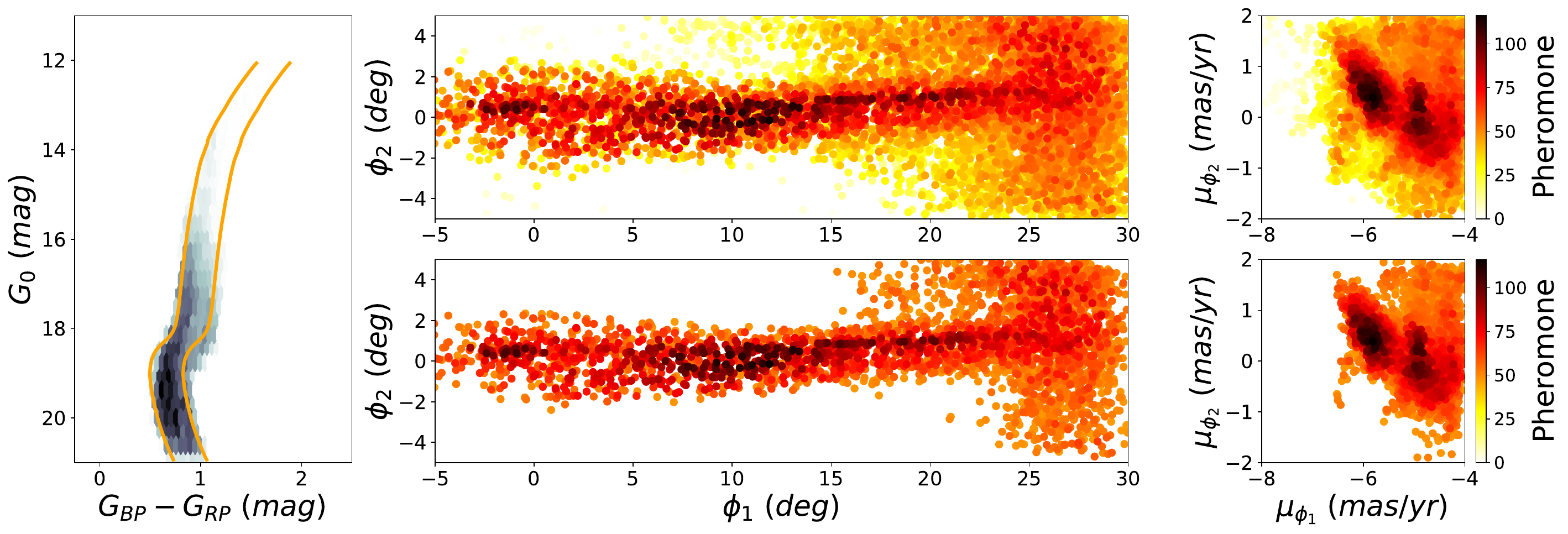}
     \caption{Second run of LAAT applied with a threshold of $40\%$.}
     \label{fig:LAAT2}
     \end{subfigure}
     \begin{subfigure}[t]{0.95\textwidth}
     \centering
     \includegraphics[width=\textwidth]{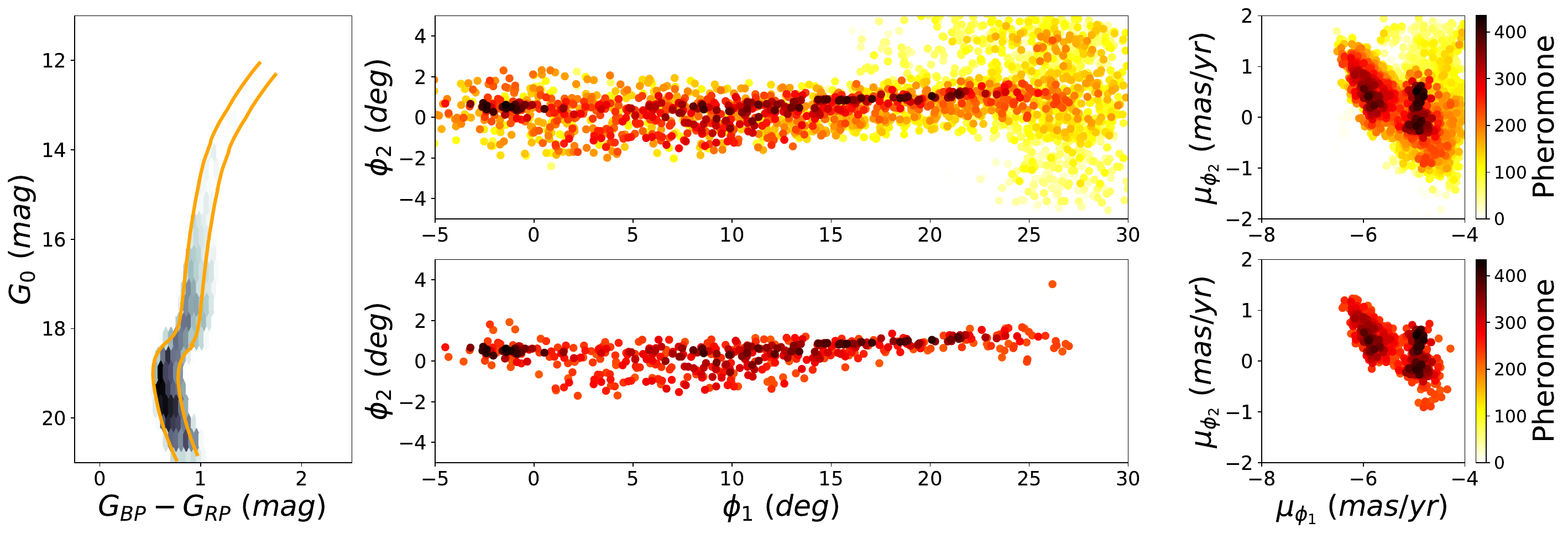}
     \caption{Third run of LAAT applied with a threshold of $50\%$.}
     \label{fig:LAAT3}
     \end{subfigure}
\caption{Three runs of LAAT (a-c) after fine-tuning the  CMD selection between each run. The CMD refinement is shown in the first column and is applied before each LAAT iteration. Middle panels show the spatial distribution of the stars while left panels show their position in proper motion space.
The upper row of the panels (a-c) shows the distribution of pheromone at the end of the run, while the lower row shows the remaining stars after enforcing a threshold on the pheromone quantity.
Darker colors mark stars that accumulated higher quantities of pheromone, indicating regions of local alignment and density. In this way, a CMD cut is first performed, followed by finding a distribution of pheromone using LAAT and retaining stars that have accumulated a pheromone quantity that exceeds the given threshold. The CMD of the remaining stars is replotted and refined and the procedure is repeated until the stream is isolated from the majority of field stars.
}
\label{fig:LAATRuns}
\end{figure*}

In the setting we have so far, the structure we would like to extract from the data is the Jhelum stream and the stars most likely to belong to it, using information of the stars' positions, proper motions and photometry. Since dynamically cold streams tend to occupy their respective proper motion spaces as local over-densities, the proper motion information of our selection of stars can thus amplify the contrast between the stars belonging and not belonging to the stream. We therefore run LAAT on the four-dimensional space $(\phi_1, \phi_2, \mu_{\phi_1}, \mu_{\phi_2})$ composed of the spatial and proper motion components of our selection of stars. We list our parameter choices used as input for LAAT and their respective definition in \tref{table:1}, and discuss them in \sref{sec:discussion}.

The procedure followed to separate the stream {members} from {contaminating nonmember} stars is the following: after defining the CMD selection in \fref{fig:PolygonSelect}, we apply LAAT to highlight the stars of interest using the pheromone deposited, and apply a cut-off threshold on the pheromone to remove as many {field} stars as possible. The remaining stars after applying this procedure are then replotted on a CMD and the polygon selection is refined so that it follows the distribution of stars more closely. The procedure is then applied iteratively in this manner, until any further selection based on the pheromone quantity would have a high chance of removing stars belonging to the stream. By getting as many reliable member stars as possible, we are able to constrain the radial velocities of each component as done in the following sections. In this way, we pinpoint the location of stream member stars in the CMD, and extract the stream from {within the data}.   

In total, we applied this procedure three times to extract Jhelum from its backgound. The CMD refinement is shown in the first column of \fref{fig:LAATRuns} where each panel refers to the three LAAT runs respectively.
The result of applying LAAT on the selection of stars described in \sref{sec:data} is displayed in the upper row of \fref{fig:LAAT1} 
where regions accumulating a larger amount of pheromone are shown in darker colors. We observe that the track of Jhelum has been prominently highlighted by the pheromone unlike regions farther away from the stream. We can also see that as we move in increasing degrees of $\phi_1$, the pheromone  quantity seems to increase on the stars of the field. This can be explained by the fact that the region on the right is closer to the Galaxy's disk and is therefore more densely populated with stars, thus, it will accumulate more pheromone. This is also seen on the right edge of the proper motion space (top-right plot of \fref{fig:LAAT1}). One can see from this result that applying LAAT once and enforcing a high pheromone cut-off instead of following an iterative procedure, will keep many contaminant stars that reside in regions of high density.
When inspecting the pheromone distribution of star members in the proper motion space, we also see the emergence of two overdensities as first noted by \citet{ShippEtal2018}. Our study is focused on the new information about Jhelum derived from these two overdensities to understand their physical meaning, and will be presented in detail in the following sections. For now however, we continue with refining our selection of stars. Once we have our pheromone distribution, we apply a threshold whereby we retain the stars that have accumulated at least $30\%$ of the maximum amount of pheromone deposited in the run. This value is chosen conservatively so as not to remove any stars that could be part of Jhelum while still removing as many contaminants as possible. The result of this step is shown in the bottom row of \fref{fig:LAAT1} where the remaining stars are plotted in both position and proper motion space. The CMD of the remaining stars is then replotted (left panel of \fref{fig:LAAT2}), and a finer selection in that space is applied such that the area within the orange outlines of \fref{fig:PolygonSelect} is shrunk. The selection is chosen so as to follow the distribution of the remaining stars in color and magnitude more closely. We then apply LAAT a second time on the refined selection, and show the results in the upper row of \fref{fig:LAAT2}. 
The application of the threshold where we keep the stars that accumulated at least $40\%$ of the maximum amount of pheromone in the run is shown in the bottom row of the same figure. We can see that the contamination from the region closest to the Galaxy's disk (right sides of the position and proper motion spaces) has been reduced while safeguarding the main structure of the stream. We also note the persistence of the two overdensities in proper motion space. We repeat this procedure a last time, where we replot the CMD of the stars remaining after the second threshold cut, fine-tune the selection region by following the distribution of the remaining stars in color and magnitude, and apply LAAT on the refined selection. The result of the third LAAT run is shown in \fref{fig:LAAT3} 
in a similar manner as for the other runs. The bottom row of the figure shows the stars that have survived a $50\%$ cut on the pheromone quantity. We observe that the contamination from field stars not likely to belong to Jhelum has been greatly reduced in this iterative application, and the track that the stream follows is more defined. From the distribution of stars in position space, we see that the pheromone accumulates more on the narrow component of Jhelum than on the broad component that extends approximately between $ 0^{\circ} \lesssim \phi_1 \lesssim 15^{\circ}$ and $\phi_2 \lesssim 0^{\circ}$ (See \citet{BonacaEtal2019} for a detailed explanations of these two components). This is expected as the narrow component is more linear and has greater directional alignment than the broad component which appears more diffuse \citep{BonacaEtal2019}. In  \tref{table:1}, we provide the properties of the three runs including the neighborhood radius $r$ that produced the clearest density contrast, the input number of stars $N_i$, the number of stars $N_f$ remaining after applying a threshold cut, as well as the time needed to perform each run.

\section{Fine-tuning and results}
\label{sec:results}

Some contamination from field stars could still be present which consists of a few remaining halo stars and stars residing closer to the Galactic disk. The latter persist in the sample due to their high local density which leads to them accumulating a large amount of pheromone. A source of contamination could be the fact that our selection is purely based on photometry, position and proper motion, as we do not know the distance or the line-of-sight velocity of the majority of these stars. The highlighting of some contaminant stars by LAAT, especially those closer to the Galactic disk, is not surprising since they reside in a dense region and will therefore accumulate large amounts of pheromone. One can eliminate those stars by increasing the threshold on the pheromone quantity. At this stage however, using a large threshold introduces a high risk of eliminating stars that are part of the stream. 

\begin{figure}
\centering
\includegraphics[width=\columnwidth]{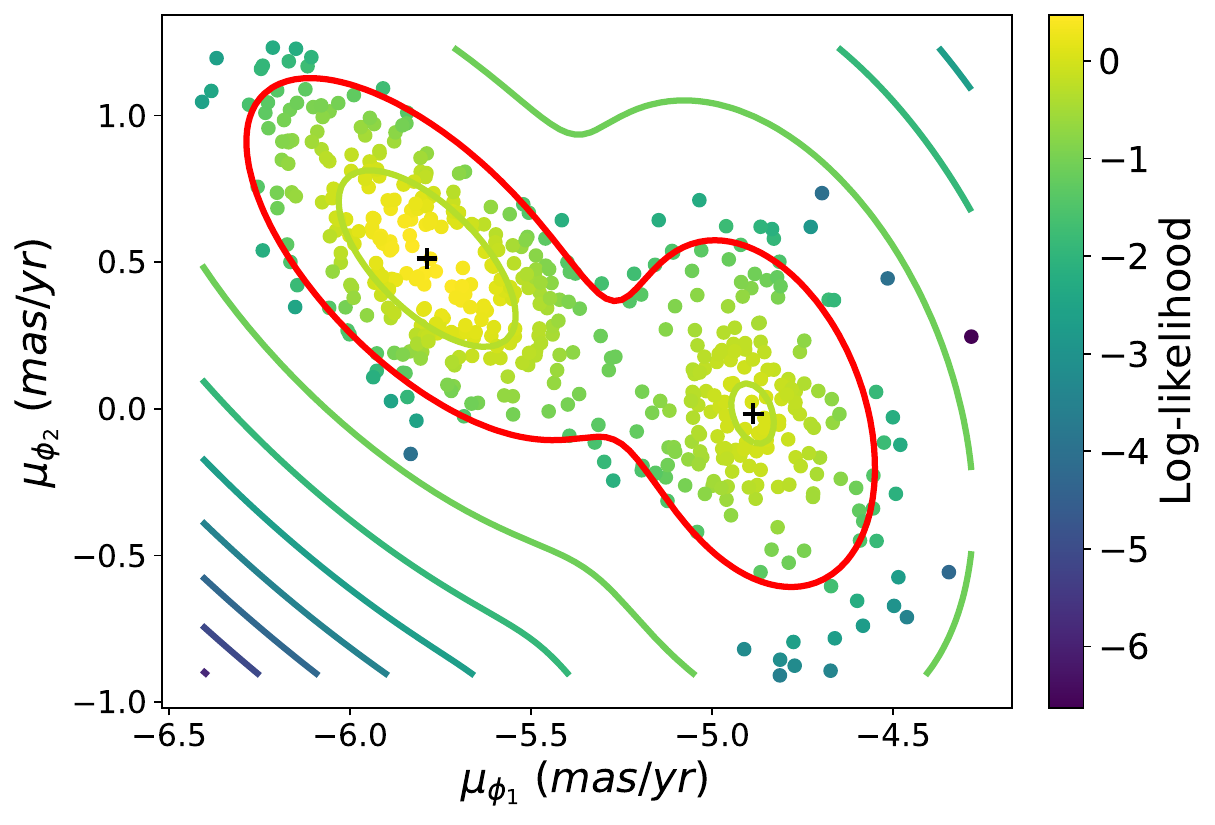}
\caption{Proper motion space of the stars remaining after the selection procedure of \sref{sec:method}. The data points within this space are attributed a score to belong to a 2-component Gaussian mixture trained on the distribution of stars in proper motion space shown in the lower left corner of \fref{fig:LAAT3}. The data points are colored by the score (log likelihood) over the star's positions in proper motion space, and contours of the log likelihood are also visualized. The contours pin-point the location and direction of the trained Gaussian distributions. We also show the centers of the two Gaussian distributions by the plus sign. The red contour corresponds to log likelihood~$>-1.75$. All data points that fall within this contour are chosen and kept for any subsequent analysis.}
\label{fig:GMMfinetune}
\end{figure}

To avoid this risk, and to make the sample as pure as possible, we use a Gaussian mixture model (GMM) to model the two overdensities in proper motion space as two\footnote{ The bottom right overdensity in proper motion space seems to have two peak signals, however upon inspection, we observed no noticeable difference between them in position space or in radial velocity. The "clumps" in that overdensity likely have a slightly different velocity because there are velocity gradients along the stream orbit.} Gaussian distributions\footnote{ { We use the GMM class provided by {\ttfamily sklearn.} Note that uncertainties on the proper motions are not taken into account, however, the median uncertainty on the proper motion measurements is 0.2~mas/yr which is five times smaller than the separation between the centers of the GMM distributions ($\approx 1$~mas/yr).}}. 
We use the proper motion of stars in the bottom right panel of \fref{fig:LAAT3} as a training set, and sample the log likelihood of these stars from the constructed mixture model over their position in proper motion space. 
The same proper motion space as in \fref{fig:LAAT3} 
but colored by the log likelihood is shown in \fref{fig:GMMfinetune}. The iso-likelihood contours are also plotted to indicate the centers and orientations of the Gaussian distributions. The means of each Gaussian is also indicated by a plus sign. This allows us to set a lower bound on the estimated likelihood to select the stars that have a better chance to belong to the model Gaussians. We enforce a lower bound of log likelihood~$>-1.75$ and keep all the stars that survive this final selection criteria, here shown as the stars within the red contour in \fref{fig:GMMfinetune}. This lower bound is chosen so as to outline the stars closer to the two overdensities in proper motion space.
The proposed modeling methodology is aimed at finding the relation between the found proper motion overdensities and their corresponding positional information. If a model of the positional information is also required, one could use existing methodologies of 1-DREAM, presented in \citet{CanducciEtal2022}. 
In the joint proper motion and positional space however, the two overdensities, may form low-dimensional manifolds of dimension larger than one. In this case, a full model for each manifold is achievable via the methodology proposed in \citet{CANDUCCI2022103579}.

\begin{figure*}
        \includegraphics[width=\textwidth]{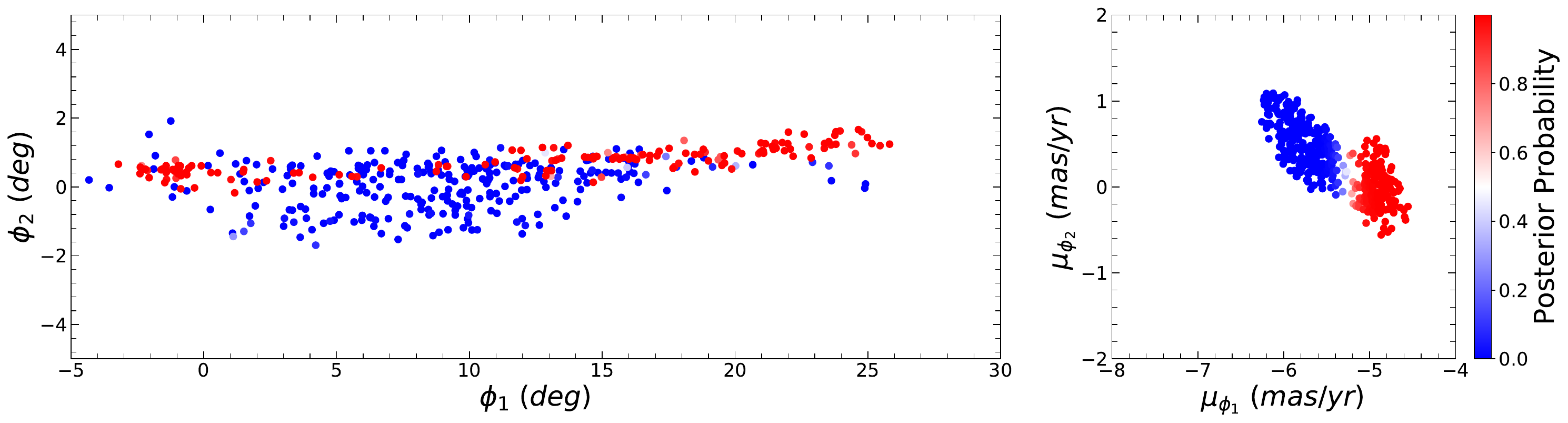}
    \caption{Posterior probabilities after clustering the two overdensities in proper motion space using a two-component Gaussian Mixture Model (GMM). Right panel: All stars in blue have a high probability of belonging to the top left overdenisty in proper motion space, while all stars in red have a high probability of belonging to the lower right overdenisty. Left panel: When visualized in position space, the two proper motion overdensities correspond to the narrow and broad components.}
    \label{fig:SeparateClusters}
\end{figure*}

Using the constructed GMM, we can calculate the posterior distributions over the two Gaussian components for each selected star. We can therefore separate the stars belonging to either overdensity accordingly. In the right panel of \fref{fig:SeparateClusters}, we show the proper motion space of our selection of stars colored by the probability density to belong to either overdensity, in this case, to the one on the bottom right. The left panel of the same figure shows where the clustered stars lie in position space. From \fref{fig:SeparateClusters}, we can see that the overdensities correspond to the narrow and broad components respectively.

We also comment on the density variations seen in our selection for the Jhelum stream components. \citet{WoudenbergEtal2023} and \citet{ViswanathanEtal2023} have reported on the presence of a tertiary component to Jhelum located on the top left side of the narrow component and is parallel to it. This component is also seen in our \fref{fig:LAATRuns} especially in the middle plots of panels (a) and (b). As this component is very faint, we see that it gets slowly filtered out with the consecutive runs of LAAT {because it is much fainter than the narrow and broad components, and so the pheromone quantity it accumulated is smaller than the threshold cuts we applied.}  We also recover the kink at $\phi_1 = 15^\circ$ similar to what is seen in \citet{ViswanathanEtal2023}. Finally, we see that the narrow and broad components are not separately parallel structures as thought to be in \citet{BonacaEtal2019}, but are in fact overlapping in their spatial distributions.

For the subsequent analysis of either component, we separate the stars within each overdensity to be able to clearly define the properties of each stream component separately. We therefore keep the stars that have at least a 50\% posterior probability to belong to either component. In total, we thus have 167 stars that we classify as belonging to the narrow component, and 279 stars that belong to the broad component of Jhelum. For the remainder of this work, we represent the stars corresponding to the narrow component in red and those corresponding to the broad component in blue. We also note that the narrow component seems un-evenly sampled, or seems much more sparse in the regions that overlap with the broad component. This however is an artifact produced by LAAT and not an intrinsic property of the narrow component. We discuss this result more thoroughly in \sref{sec:discussion}. 

\begin{figure}
\centering
\includegraphics[width=\linewidth
]{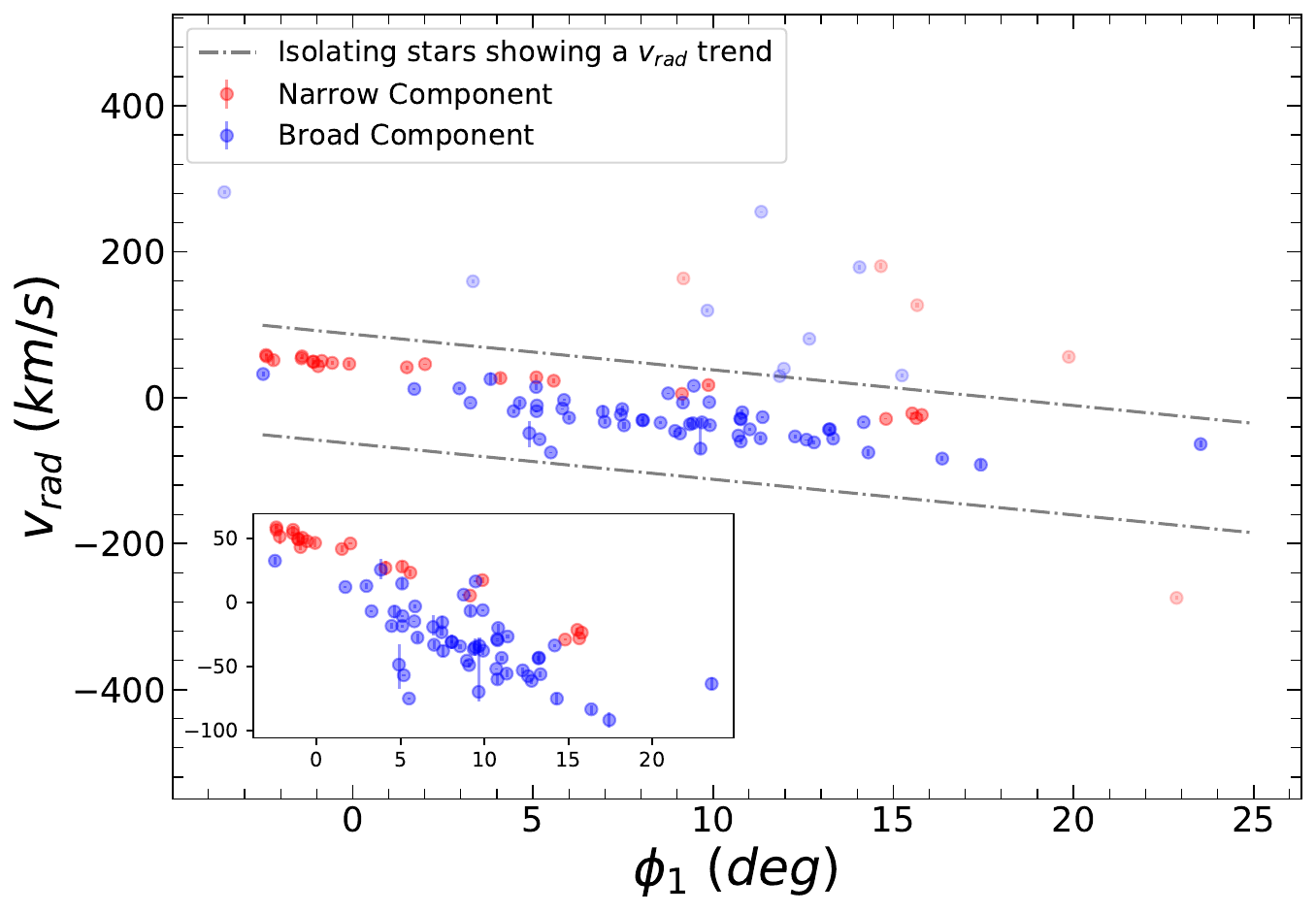}
\caption{ $v_{rad}$ versus $\phi_1$ plot for the separated stars between the narrow and broad components. The radial velocity is obtained from measurements by the $S^5$ survey. When isolating the stars that show a trend in radial velocity i.e., those included in the dashed gray lines, we observe a separation between the two components in radial velocity. A zoom-in plot of the considered region is plotted in the bottom left corner. Stars belonging to the narrow component additionally show a narrow spread in radial velocity compared to the more dispersed broad component.}
\label{fig:VradSeparate}
\end{figure}

We perform a final check to see if the separation between the two components also exists in the third velocity component, as in the radial velocity. For that, we use the high precision radial velocity measurements from the $S^5$ survey.
We thus plot the radial velocity as a function of $\phi_1$ of the stars for which we find radial velocity measurements (30 and 65 stars for the narrow and broad components respectively). Particularly, we focus on the stars which show a trend in terms of radial velocity outlined by the dashed gray lines in \fref{fig:VradSeparate}, and reproduced in the zoom-in plot within the figure. We observe for the first time, an offset between the narrow and broad components in radial velocity. The trend of the narrow component is composed of 22 stars while that of the broad component is composed of 54 stars. We provide the list of this star selection and its properties in Appendix~\ref{sec:appendixC}. We also see a wide radial velocity dispersion in the broad component as opposed to a narrower one for the narrow component. This has not been observed before and confirms the separation of the narrow and broad components in all three velocity components. The rest of this work focuses on extracting information from this finding and discussing the possible progenitors and formation scenarios that could have formed the stream Jhelum.

\section{Narrow and broad component properties}
\label{sec:properties}

In this section, we explore the properties of the narrow and broad component of Jhelum, separated using the procedure explained in Sections~\ref{sec:method} and~\ref{sec:results}. In particular, we attempt to fit an orbit that models the dynamics of the components in \sref{sec:orbit}, and examine the velocity dispersion, the width, and metallicity dispersions of both components in Sections~\ref{sec:veldispersion} to ~\ref{sec:metaldispersion}. The size of the dispersions provides pieces of evidence toward the type of progenitor that has formed the Jhelum stream and/or its subsequent evolution.

\subsection{Best-fit orbit}
\label{sec:orbit}

We now determine the orbits which follow the track of both the narrow and broad components of Jhelum when integrated in an axi-symmetric Milky Way potential. {Through fitting the orbits we also} estimate the dynamical properties of the components including their {widths and} velocity dispersions.  We follow the procedure thoroughly detailed in \citet{WoudenbergEtal2023} which is recounted here. To set up the Milky Way gravitational potential, we follow \citet{Price-WhelanEtal2020} and create a composite model consisting of a bulge, disk, and dark matter halo. Similar to \citet{WoudenbergEtal2023}, the bulge potential is modeled as a Hernquist sphere \citep{Hernquist1990} with a mass of $4\times 10^9 M_{\odot}$ and scale length of $c_b=1$~kpc. We model the disk as a Miyamoto-Nagai potential 
\citep{MiyamotoNagai1975} with a mass of $5.5 \times 10^{10} M_{\odot}$, scale length $a_d= 3$~kpc, and scale height $b_d=0.28$~kpc. Finally, the dark matter halo is modeled as a generalized Navarro-Frenk-White (NFW) potential \citep{NavarroFrenkWhite1996} with a mass of $0.7\times 10^{12}M_{\odot}$, scale radius $r_s = 15.62$~kpc, and a minor-to-major axis ratio $q_z=0.95$.


Any orbit integration is performed using the package {\ttfamily AGAMA} \citep{Vasiliev2019} with the above defined potential. The integration of the stars' orbits is performed in Galactocentric coordinates. We are fitting a single orbit to each component although the components have a given width or dispersion, as the stars belonging to the components do not all follow the exact same orbit. The single orbit fit however, is a good approximation of the best fit (see Appendix A in \citet{WoudenbergEtal2023}) and will therefore be used here. With this information, we use the Markov chain Monte Carlo (MCMC) method to find the model parameters that best fit the data. The orbit model parameters consist of the declination $\delta$, distance to the stream $D$, the proper motion components $\mu_{\alpha}$ and $\mu_{\delta}$, as well as the radial velocity $v_{rad}$. The right ascension $\alpha$ on the other hand is kept fixed throughout the run to avoid degenerate solutions to the best fit orbit. We measure the fitness of an orbit in following the track of the set of stars by defining the log-likelihood function ${\rm ln}(L)$: 

\begin{equation}
    {\rm ln}(L) = \frac{1}{N} \sum_{i=1}^N{
    \left[-{\rm ln}
    \left(\prod_{j\in \chi}{(2\pi)^{\frac{1}{2}
    }\sigma_{ij}}\right)-
    \frac{1}{2}\sum_{j \in \chi}{\left(\frac{x^d_{ij}-x^m_{ij}}{\sigma_{ij}}\right)^2}
    \right]} \; .
    \label{eq:loglike}
\end{equation}

The index $j \in \chi$ where $\chi = \{ \delta, \mu_{\alpha}, \mu_{\delta}, v_{rad}\}$, and so $x_{ij}$ refers to the $j$-th quantity of the $i$-th star in the data. The superscript $m$ refers to the modeled orbit evaluated on the location of the data points while the superscript $d$ refers to the data measurements. Moreover, $\sigma$ denotes the errors on the four quantities we attempt to fit. The errors consist of both the measurement errors $\sigma_{meas}$ and the intrinsic dispersion of these quantities along the stream {$\sigma_{int} \in \{ \sigma_w, \sigma_{\mu_{\alpha}}, \sigma_{\mu_{\delta}}, \sigma_v \}$ where $\sigma_w$ is the component width, $\sigma_{\mu_{\alpha}}$ and $\sigma_{\mu_{\delta}}$ are the transverse velocity dispersions and $\sigma_v$ is the radial velocity dispersion}. The error in \eref{eq:loglike} is therefore calculated as the sum in quadrature of these two uncertainties: $\sigma = \sqrt{\sigma_{meas}^2 + \sigma_{int}^2}$. $\sigma_{ij}$ therefore refers to the error on the $j$-th quantity for the $i$-th data point. {The intrinsic dispersions are kept as free parameters that we fit along with the orbit modeling parameters.} 

An initial guess for the model parameters is taken as the measured values of a randomly chosen star belonging to the corresponding component, and the right ascension $\alpha$ is fixed to the value corresponding to the chosen star. For the {parameters we are fitting}, we set a flat prior defined by the following:
\begin{equation}
    P(x)=
    \begin{cases}
        \; 1 & \text{if} \begin{cases}
            \quad -53^{\circ} < \delta < -47^{\circ}\\
            \quad 0  < d/\text{kpc} < 20\\
            \quad 5  < \mu_{\alpha}/\text{mas yr}^{-1} < 8\\
            \quad -7  < \mu_{\delta}/\text{mas yr}^{-1} < -3\\
            \quad -125< v_{rad}/\text{km s}^{-1} < 60 \\
            \quad 0^{\circ} < \sigma_w < 10^{\circ} \\
            \quad 0 < \sigma_{\mu_{\alpha}} /\text{mas yr}^{-1} < 10 \\
            \quad 0 < \sigma_{\mu_{\delta}} /\text{mas yr}^{-1} < 10 \\
            \quad 0 < \sigma_w /\text{km s}^{-1} < 30 \\
        \end{cases}\\
        \; 0 & \text{otherwise.}
    \end{cases}
\end{equation}
In this way, the MCMC algorithm is run using the {\ttfamily emcee} package with 80 walkers and 1000 steps to ensure convergence.

\begin{figure*}
\centering
    \includegraphics[width=\textwidth]{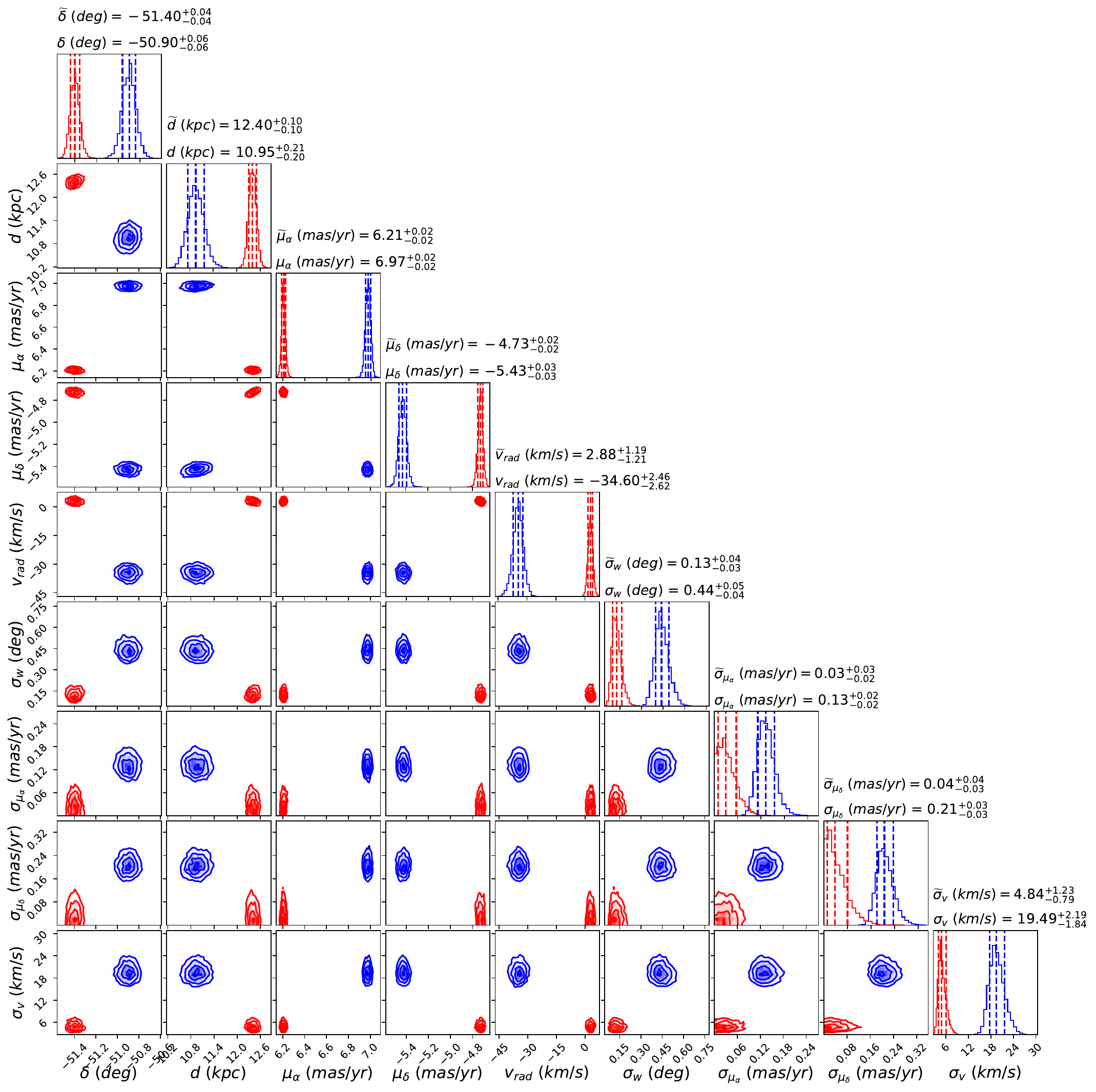}
\caption{Posterior distributions for all parameters modeled using a Markov chain Monte Carlo (MCMC) algorithm to obtain a best-fit orbit for the narrow (red) and broad (blue) components of Jhelum as well as the component widths and velocity dispersions. The median of each modeled parameter is indicated on top of each column along with the $16$-th and $84$-th percentile variations. Values indicated with or without a tilde on top of each column refer to the narrow or broad component, respectively.}
\label{fig:Cornerplots}
\end{figure*}

\begin{figure}
\centering
    \includegraphics[width=\linewidth
    ]{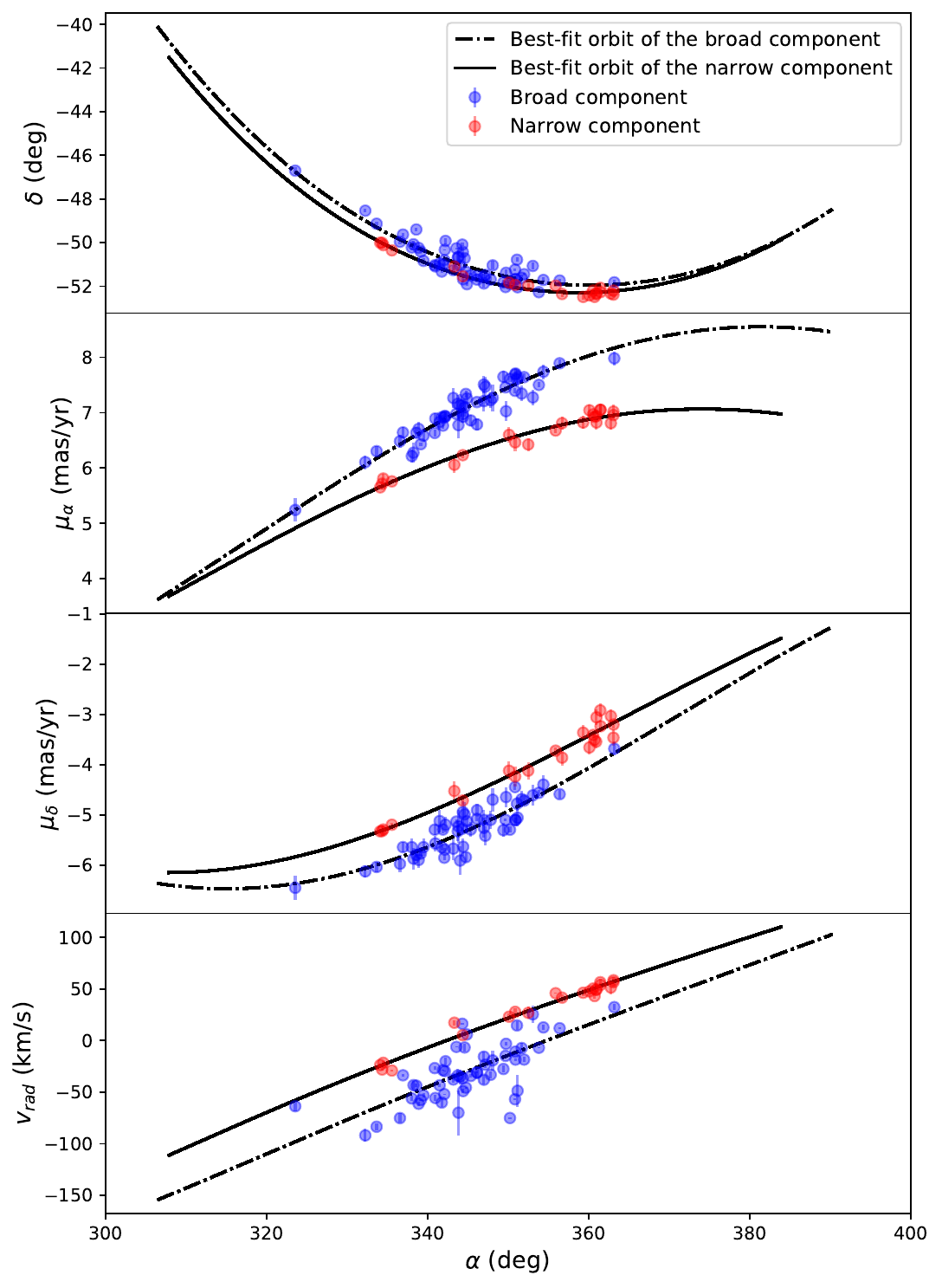}

\caption{Best fit orbits for the narrow and broad components of Jhelum in a standard Milky Way potential. Stars belonging to the narrow and broad component that were observed by the $S^5$ survey are shown in red and blue, respectively. The black solid and dashed lines in both panels indicate the best fit of orbit of each component respectively.}
\label{fig:BestFitOrbit}
\end{figure}

\begin{figure}
    \centering
    \begin{subfigure}[t]{\linewidth}
    \centering
    \includegraphics[width=\linewidth]{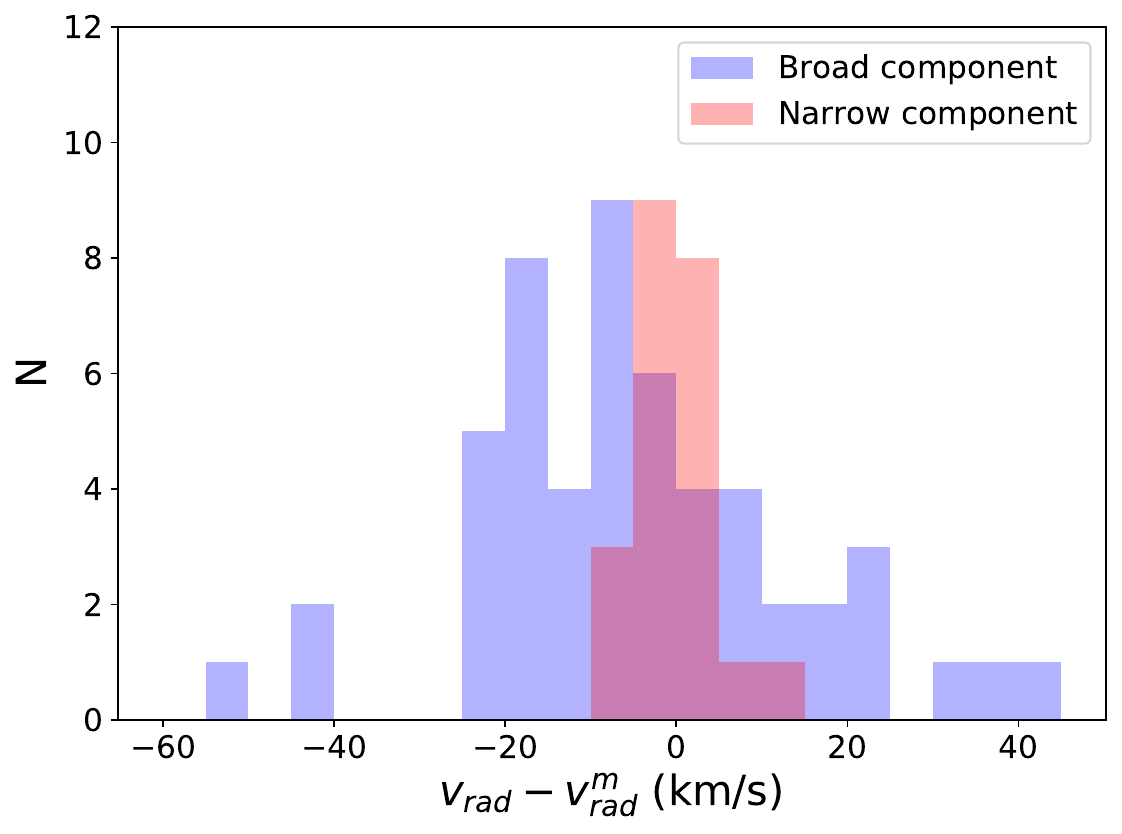}
    \label{fig:veldisp}
    \end{subfigure}
    \begin{subfigure}[t]{\linewidth
    }
    \centering
    \includegraphics[width=\linewidth]{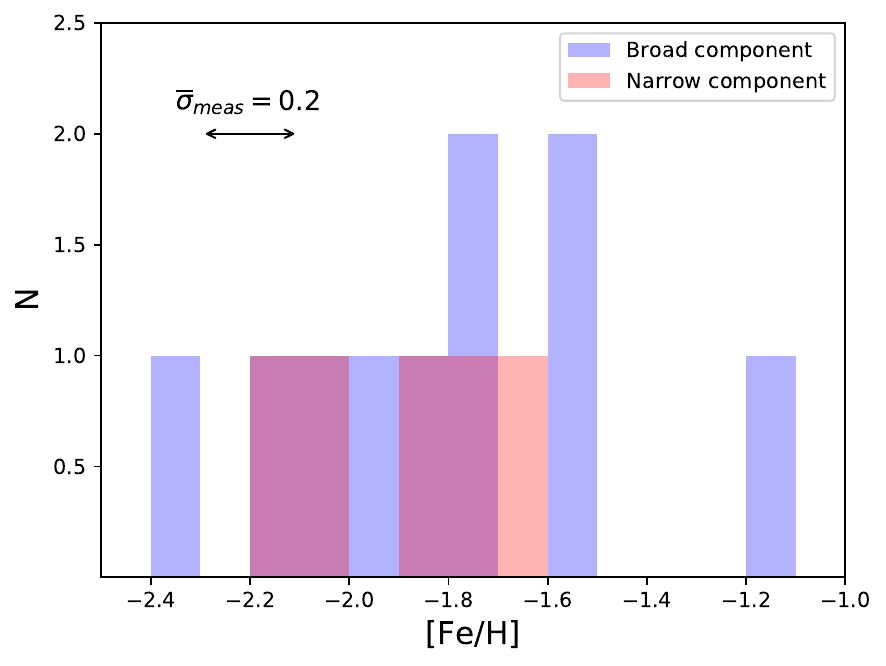}
    \label{fig:Feh}
    \end{subfigure}

    \caption{Radial velocity and metalicity distributions for either components of Jhelum. Top panel: Distribution of the radial velocities around the best-fit orbit of each component of the stream. The width of these distributions {helps us visualize} the velocity dispersion $\sigma_v$ for the broad component, and $\tilde{\sigma}_v$ for the narrow component. Lower panel: Distribution of metallicities, [Fe/H], for either component of the stream. Metallicity measurements are obtained from the $S^5$ survey. The intrinsic width of these distributions is fitted to a Gaussian distribution, to find the metallicity dispersion of each component. This calculation is explained in \sref{sec:metaldispersion}. We also indicate the mean error on the metallicity measurements, $\overline{\sigma}_{meas}=0.2$.}
    \label{fig:histograms}
\end{figure}


The corner plot of the posterior distributions of all modeled parameters is shown in \fref{fig:Cornerplots} (red and blue corresponding to the narrow and broad components respectively). The 50-th percentile (median) values for each fitted quantity is also portrayed in the figure with the errors given by the 16-th and 84-th percentiles (values indicated with a tilde correspond to the narrow component). The median values are taken to be the parameters that produce the best fitting orbits for the stream's components. Regarding the narrow component, we see a strong degeneracy between the distance parameter $d$ and $\mu_{\delta}$ and weaker degeneracies between combinations of the other quantities. This degeneracy is also weaker for the distributions of the broad component which also shows wider {posterior} distributions. One can explain this given the fact that the broad component is shorter and more diffuse than the narrow component. Therefore, a wider range of parameters produces orbits that fit the distribution of stars well, that is, the broad component data is less constraining. In terms of the distance to the stream, we retrieve a median value of $12.40^{+0.10}_{-0.10}$~kpc for the narrow component and $10.95^{+0.21}_{-0.20}$~kpc for the broad component. 
{These distance measurements agree with recent estimates from works such as \citet{LiEtal2022}, \citet{WoudenbergEtal2023} and \citet{ViswanathanEtal2023}, but are nonoverlapping given their respective error margins. The difference between the distances to the two components is $\approx 1.4$~kpc. We discuss this result further in Section~\ref{sec:discussion}}. 
In \fref{fig:BestFitOrbit}, we provide the best fit orbit of the narrow and broad components plotted using solid and dashed black lines respectively. The stars belonging to either component are shown in their respective colors. We see that the orbits fit the distribution of most of the stars of either component well. 

\subsection{Velocity dispersion}
\label{sec:veldispersion}
{Through this orbit fitting, we also obtain estimates of the velocity dispersions and widths of Jhelum's narrow and broad components.} Given $v_{rad}$ as the radial velocity measurements from $S^5$ and $v^m_{rad}$ as the radial velocity obtained from the best-fit orbit, {the radial velocity dispersion can then be visualized as the width of the distribution of $v_{rad} - v^m_{rad}$ as shown in the top panel of \fref{fig:histograms}. From the MCMC fit, the radial velocity dispersions are found to be equal to $4.84^{+1.23}_{-0.79}$~km/s and $19.49^{+2.19}_{-1.84}$~km/s respectively. As for the dispersions in $\mu_{\alpha}$ and $\mu_{\delta}$, we obtain $0.03^{+0.03}_{-0.02}$~mas/yr ($1.75^{+1.78}_{-1.17}$~km/s) and $0.04^{+0.04}_{-0.03}$~mas/yr ($2.33^{+2.37}_{-1.75}$~km/s) respectively for the narrow component, and $0.13^{+0.02}_{-0.02}$~mas/yr ($6.69^{+1.18}_{-1.13}$) and $0.21^{+0.03}_{-0.03}$~mas/yr ($10.81^{+1.78}_{-1.71}$~km/s) respectively for the broad component. The conversion of units has been performed assuming the best-fit distance obtained for each component. The radial velocity dispersion will be referred to as the velocity dispersion hereafter and the} implication of these measurements will be discussed in \sref{sec:discussion}.

\subsection{Stream width}
\label{sec:width}

The width of a stream can also provide information on its progenitor. Wider streams tend to be a result of a DG falling into the Milky Way potential, while GC accretion tends to produce narrower streams. Since we have a distinct sample of high confidence members for the narrow and broad 
component of Jhelum, we estimate the width of either of these components separately
rather than calculating one width for the entire stream. {The component width $\sigma_w$ is calculated as described in the previous subsection when fitting for the orbit of either component of the stream.} With this procedure, we obtain $\sigma_w \sim 0.13^{\circ}$ for the narrow component of Jhelum and $\sigma_w \sim 0.44^{\circ}$ for the broad component. Assuming the posterior best-fit distances shown in \fref{fig:Cornerplots} ($12.40^{+0.10}_{-0.10}$ and $10.95^{+0.21}_{-0.20}$), we find the linear widths of the components. The evaluated widths are then $\sigma_w = 28.13^{+8.9}_{-6.64}$~pc and $\sigma_w = 84.09^{+11.26}_{-7.17}$~pc. We provide a comparison between these estimates and those calculated by other works such as {\citet{BonacaEtal2019} and} \citet{ShippEtal2019} in \sref{sec:discussion}.

\subsection{Metallicity dispersion}
\label{sec:metaldispersion}

For evaluating the mean metallicity, $\overline{[Fe/H]}$, and metallicity dispersion, $\sigma_{\rm [Fe/H]}$, of either component of the stream, we utilize the metallicities provided by the $S^5$ Survey calculated using the Calcium Triplet (CaT) regions. The CaT metallicities have been derived for red giant branch stars (RGB) of the stream using the equivalent widths (EW) of the CaT lines and using the EW to metallicity calibration from \citet{CarreraEtal2013}. From these measurements, the $S^5$ Collaboration has provided us with high-quality members stars as found in \citet{LiEtal2022} which we then use. These measurements were reported as the more trusted estimates in \citet{LiEtal2022} and have been used as the basis of the discussion around the chemical properties of the dozen streams studied within the same work. With our selection of stars that belong to Jhelum, 15 stars have their metallicities measured by the survey. This sample is much smaller than the radial velocity sample and the measurements are mostly for stars of the broad component. Of these measurements ten are for stars belonging to the broad component and five are for stars in the narrow one. We display the distribution of metallicities for each component in the lower panel of \fref{fig:histograms}.

We then run an MCMC algorithm to model $\overline{[Fe/H]}$ and $\sigma_{\rm [Fe/H]}$ of either component by fitting a Gaussian function to the distribution of metallicities of each component. Similar to \sref{sec:orbit}, the total width $\sigma$ of the metallicity distribution of a given component is given by the sum in quadrature of the measurement errors and the distribution's intrinsic width which we are attempting to fit. Therefore, we have $\sigma = \sqrt{\sigma_{meas}^2 + \sigma_{int}^2}$. The initial guess for the MCMC algorithm is taken as mean and standard deviation of the selection of metallicity measurements we have for either component. The fit is then performed by optimizing the following log-likelihood function: 

\begin{equation}
    {\rm ln}(L) = \sum_{i=1}^N {\rm ln}
    \left( \frac{1}{\sqrt{2\pi}\sigma_i}{
    \exp}
    \left[-\frac{1}{2} 
    \left(\frac{[Fe/H]_i - \overline{[Fe/H]}}{\sigma_i} \right)^2 
    \right] \; \right)\; .
    \label{eq:loglike2}
\end{equation}

\begin{figure}
\centering
\includegraphics[width=\linewidth
]{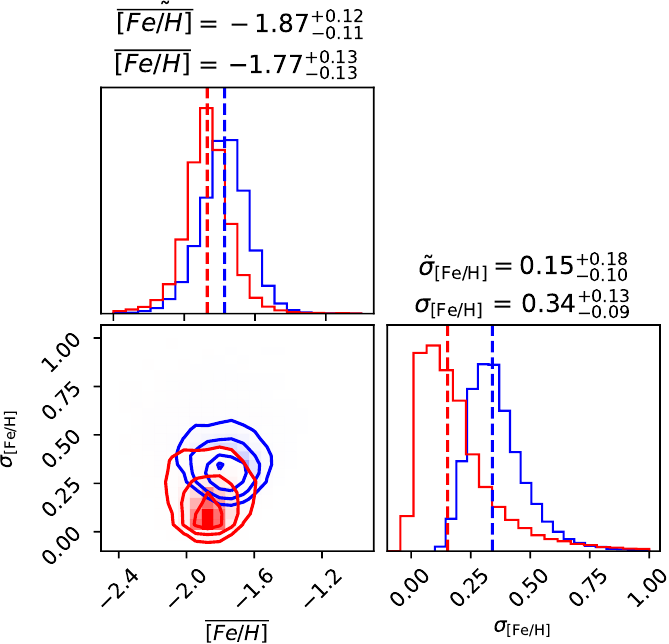}
\caption{Posterior distributions of the mean metallicity $\overline{[Fe/H]}$ and intrinsic metallicity dispersion $\sigma_{int}$ modeled using an MCMC algorithm to obtain a best-fit Gaussian for the distribution in the lower panel of \fref{fig:histograms}. Values indicated by a tilde refer to the narrow component.}
\label{fig:MetallicityCorner}
\end{figure}

We run the MCMC algorithm with 80 runners and 1000 steps and then extract the best-fit parameters as the 50-th percentile values of the resulting distribution. The posterior distributions for the Gaussian fits is shown in \fref{fig:MetallicityCorner} where we also display the best fit modeled mean metallicities and metallicity dispersions. We denote $\sigma_{int}$ for the broad component as $\sigma_{\rm [Fe/H]}$ and that for the narrow component as $\tilde{\sigma}_{\rm [Fe/H]}$. For the narrow component, we obtain a mean $\tilde{\overline{[Fe/H]}}= -1.87^{+0.12}_{-0.11}$ and a metallicity dispersion of $\tilde{\sigma}_{\rm [Fe/H]} = 0.15^{+0.18}_{-0.10}$, while for the broad component we obtain $\overline{[Fe/H]}= -1.77^{+0.13}_{-0.13}$ and $\sigma_{\rm [Fe/H]} = 0.34^{+0.13}_{-0.09}$. Note that for the narrow component, the posterior distribution intersects with $\sigma_{int} = 0$, and given the low amount of stars with metallicity measurements for this component, the metallicity dispersion represents an upper bound of its actual dispersion. The comparison of the calculated values with other estimates in literature is performed in \sref{sec:discussion}.

\section{Discussion}
\label{sec:discussion}

In this section, we discuss the procedure and results explained throughout this work. We review the robustness of the selection criteria of the stars belonging to Jhelum and the dependence of our results on the used methodology. We also discuss the different formation scenarios of the stream Jhelum based on the results achieved in \sref{sec:properties}.

\subsection{Evaluation of selection procedure}

The selection of stars depending on their position in the CMD is a standard step followed to isolate those that are members of a given stellar population. In \sref{sec:data} we defined the initial polygon selection to be as wide as possible to ensure that all stars belonging to Jhelum fall within this region, even though this also includes many of the surrounding field stars. Given that LAAT, in its current version, uses one value of pheromone threshold to filter out stars within an entire run (as opposed to using a threshold which is dependent on the location within the distribution of stars) it is possible to miss some stars that are true members of the stream. Therefore, since several filtering procedures follow this initial selection (see \fref{fig:LAATRuns}), it becomes necessary to be as inclusive as possible in the first step of the CMD selection. This fact becomes important when considering that much less stars occupy the red giant branch (RGB) than the main sequence part of the CMD. Thus, missing some stars that belong to this branch limits the subsequent metallicity dispersion analysis that depends on the high quality measurements of these stars. This risk is also mitigated by the iterative approach we follow whereby the CMD selection is repetitively narrowed down, guided by the distribution of the remaining stars after each run of LAAT. 

The parameters chosen for running LAAT are listed in \tref{table:1} and here we explain the intuition behind choosing the values for these parameters. The large number of agents and number of steps have been chosen to make sure that each star has been visited multiple times during the run. This allows for the convergence of the algorithm toward a result that does not change between different initializations of the random walk. The neighborhood radius parameter $r$ defines the region in which PCA is performed to determine the main orthogonal directions along which the stars are distributed. If $r$ is smaller than necessary, then the stars within the neighborhoods may be insufficient to infer any alignment information. This parameter is thus chosen such that it is large enough to create a region where the linearity of Jhelum is detected. Similarly, if $r$ is larger than necessary, many field stars would be included in the neighborhood which could drown the alignment signal. A large neighborhood radius also acts as a zoomed out perspective of the region it encompasses, and so leads to missing out on smaller structures in the data that are distributed on a smaller scale. For example, a large radius will lead to highlighting the two clusters in proper motion space as one large overdensity. Note however that LAAT does not create false-positive detections of structure, but rather highlights local density contrasts (especially if aligned along a preferred subspace) that would not have been seen on larger scales (for a detailed proof, see \citet{LAAT, Mohammadi2022}). In other words, the separation of the clusters in proper motion space is not an artificial creation of the algorithm. This claim is substantiated by the fact that the two overdensities show a correspondence to the narrow and broad component (see \fref{fig:SeparateClusters}) and by the separation present in radial velocity as well (see \fref{fig:VradSeparate}). As for the remaining parameters, they have not been altered from the default settings of the algorithm.

The choice of $r$ in this work has been picked to be on the order of the width of the stream so as to capture the most directional information, and the specific values are chosen so as to produce the highest density contrast visible. {This has been achieved by experimenting with different values for the radius and observing that $r = 0.5$ gives the best results. After performing the pheromone cut on the first run, some sparse neighborhoods will be formed in places were the density is low. To infer the main directions the stars are distributed within a neighborhood of size $r$, LAAT needs a minimum of four stars within that neighborhood if applied to four-dimensional data. For the second and third runs, since some sparse distributions form due to the applied pheromone cut, we increase the value of the radius to insure that this condition is met. The specific values again are chosen through manual experimentation and visually checking what produces the largest density contrast between the stream and field stars.}

{As for the choice of the pheromone threshold for each run, the value is first chosen conservatively to avoid eliminating member stars by mistake. Any value smaller than $30\%$ would unnecessarily keep some of the field stars that have accumulated a very small pheromone amount. If we use this same threshold value for the rest of the runs, we would retain more contaminant stars with each run which would necessitate performing more iterations of the CMD fine-tuning and running LAAT. Therefore, to keep the number of iterations within 3 runs and to remove as many nonmember stars as possible, we increase the pheromone threshold gradually from $30\%$ to $50\%$. Through experimentation, we find that smaller thresholds will keep more field stars, and larger ones would remove parts of the two components of the stream. That is why after the third run, we rely on the GMM log-likelihood cut in Section~\ref{sec:results} to fine-tune this selection instead of using harsher pheromone thresholds. We also provide further discussion along the lines of contamination and completeness of our sample in Appendix~\ref{sec:completeness}.}

We also discuss the sparsity of the narrow component in the regions overlapping with the broad component of the stream. The broad component of Jhelum is sparser than its counterpart, and the overlapping region between the two components has a relatively smaller alignment between its member stars than regions belonging solely to the narrow component. These two reasons lead LAAT to see a smaller contrast between the overlapping region of the two components and the field stars. Therefore, the broad component along with the overlapping region will acquire a low pheromone concentration compared to locations occupied by the narrow component alone. When applying the threshold on the pheromone quantity, some of the member stars that did not receive a large enough pheromone concentration will get filtered out. The part of the narrow component in that region will then be less populated as a result of the enforced cuts and will appear disconnected in some places as seen in \fref{fig:SeparateClusters}. Note that the thresholding criteria of LAAT are being updated in future versions of the algorithm {so that overlapping streams could be extracted in a more efficient manner. The idea is to implement local thresholds depending on the pheromone quantity of local neighborhoods in the data rather than using one global threshold on the pheromone for the entire input dataset. In this way, regions that show smaller alignment or smaller density but are still equally interesting will not be filtered out as harshly. Given that this is a future implementation,} with current means, we prefer to create a high purity sample at the cost of missing some member stars over creating a sample that could contain some contamination from stars in the field.

\subsection{Likely merger scenarios}

The properties estimated in \sref{sec:properties} of the component widths and velocity/metallicity dispersions are pieces of information that point at the nature of the past progenitor of Jhelum. For the narrow component of the stream, we measure a velocity dispersion of 4.84~km/s and an upper limit to the metallicity dispersion of $0.15$ dex. We also obtain a component width of $\approx 28$~pc for the narrow component. The velocity dispersion of this component is comparable to other streams studied in the literature which are classified as having a GC origin. Some of these streams are 300S \citep{FuEtal2018}, Willka Yaku, Jet, and Phoenix \citep{ShippEtal2018}, as well as GD-1 \citep{GiallucaEtal2021}. GCs are also characterized by a negligible metallicity dispersion not greater than 0.05~dex especially compared to larger systems such as DGs, which usually have metallicity dispersions an order of magnitude higher. The calculated metallicity dispersion of the narrow component is an upper limit to the true dispersion of this component's progenitor. This upper limit is again comparable to the dispersions of the streams in \citet{LiEtal2022} that are more likely results of GC accretion. Therefore, the thin and dynamically cold nature of this component, suggest that a GC might have been its progenitor. The properties of the narrow component also rule out some other origin scenarios that were hypothesized in \citet{BonacaEtal2019} particularly Jhelum being the result of multiple orbital wraps. This scenario is now questionable since it is unlikely that the narrow component would remain of such a small width after long periods of orbit in the Galaxy's potential.


As for the broad component, we measure a velocity dispersion of 19.49~km/s as well as a metallicity dispersion of $0.34^{+0.13}_{-0.09}$~dex and component width of $\approx 84$~pc {located at a distance of 1.45~kpc closer than the narrow component}. Such large dispersions can be explained by dynamical perturbations \citep{WoudenbergEtal2023}, or could be indicative of a DG origin. The latter {complements} the fact that when considering the whole stream, Jhelum has been classified so far as being more likely a remnant of a DG \citep{JiEtal2020, BonacaEtal2021, LiEtal2022}. This shows that when studying the Jhelum stream, it is important to treat the two components distinctly as highlighted by this work, otherwise one risks artificially inflating the dispersions for the whole object.
Using the updated mass-metallicity relation from \citet{RomeroGomezEtal2023}, the mean metallicity of the broad component allows us to infer the stellar mass of the progenitor to be between $10^6$ and $10^7$~$M_{\odot}$.
Our estimate of the stream width using the procedure defined in \sref{sec:width} is smaller than the value reported in works such as \citet{ShippEtal2018} and cited in \citet{LiEtal2022}. One possible cause of the dissimilarity is the difference of the processes used to measure the stream width. \citet{ShippEtal2018} attempt to fit the transverse stream profile with a Gaussian stream model and a linear foreground component. The separation between member and {contaminating} stars is performed by iteratively narrowing down their star selection around the best-fit isochrone of the CMD. On the other hand, we {fit each component width as a free parameter in our MCMC scheme.} We also consider the proper motion of the stars and their color and magnitude information to determine their membership to the stream, whereas \citet{ShippEtal2018} use photometry only. Furthermore, using LAAT, membership to the stream is determined by assigning a global threshold to all stars in a run. {Since central parts of streams tend to be more populated than their outer parts, LAAT will concentrate more pheromone on those inner regions, and so when applying a cut on the pheromone value, it is possible that some stars on the edges of the streams are filtered out while preserving the inner denser parts.} This creates the possibility of missing some member stars especially in less directionally alligned or {in the diffuse outer regions of streams}. As a result, our {calculations} could underestimate the stream width. {Furthermore, Bonaca et al. (2019) estimate the widths of the components to be $91^{+4}_{-13}$~pc and $213^{+8}_{-23}$~pc for the narrow and broad components respectively, at an assumed distance of 13~kpc. These estimates again are larger than our measurements of the component widths ($28.13^{+8.9}_{-6.64}$~pc and $84.09^{+11.26}_{-7.17}$~pc respectively). Similar to the explanation above, we attribute this difference to the fact that LAAT uses a global threshold on the pheromone quantity. We prefer to use the threshold values mentioned in this work to limit contamination as much as possible and allow better constraints on the dynamical and metallicity properties of the components, keeping in mind that it comes at the expense of underestimating the widths of the stream components.} 

As a counterargument to the possible GC progenitor of the narrow component, works such as \citet{WalkerEtal2007} and \citet{MinorEtal2010} also quote a range of 
$4-10$~km/s for the velocity dispersion of several dwarf spheroidal galaxies around the Milky Way. Moreover, the errors we have on the metallicity dispersion measurements are large especially given the relatively low amount of stars that were available to perform the measurement. Therefore, even though our measurements favor a GC accretion scenario, it is difficult to completely rule out a DG origin. On the other hand, a DG with a velocity dispersion of $\sim5$~km/s should have a stellar mass of about $10^{4-5}~M_{\odot}$ when using the relation between velocity dispersion and stellar mass from \citet{EftekhariEtal2022}. This means that the number of stars in this component should be at least a factor 100 smaller than the number of stars in the broad component. This shows that although a DG progenitor for the narrow component cannot be ruled out completely, it remains unlikely. The thin nature of the narrow component could also point toward a scenario where a nuclear star cluster (NSC) at the center of a DG has been accreted onto the Milky Way (see \citet{NeumayerEtal2020} for a review on NSCs). However, given that the narrow component is located at the edge of the broad component and not the center, it would be difficult to argue for this scenario without more data and/or modeling.

The work of \citet{WoudenbergEtal2023} has also investigated the effects induced by encounters between Jhelum and the Sagittarius DG. Given that Sagittarius and Jhelum share the same orbital plane, it is natural to assume that Sagittarius has induced dynamical and structural perturbations on the smaller stream. Through N-body simulations of a loose GC set on Jhelum's orbit, and integrating the orbits of the large perturber and that of the GC in Milky Way-like potentials, \citet{WoudenbergEtal2023} have shown that encounters between the two systems produces multiple components in Jhelum’s stream. Their simulations also show that the interactions with Sagittarius result in inflating the measured velocity dispersion of Jhelum by a factor of 4 at most, compared to the unperturbed stream. These interactions are also elements that explain Jhelum's complex morphology. \cite{WoudenbergEtal2023} also point out a tertiary component of Jhelum located on the top left of the narrow component and is parallel to it. The component can be seen in \fref{fig:LAATRuns} but is slowly filtered out upon applying the multiple runs of LAAT, and so we leave its exploration for future work.

Given this information, our measurements show a likely multiple progenitor scenario of Jhelum in which a GC belonging to a DG was accreted onto the Milky Way during the DG's infall, and produced the remnants that form the Jhelum stream. Arguments of this kind are also present in \citet{ErraniEtal2022} when discussing the possible progenitor of the C-19 stream.{The difference in the distance to each component would correspond to an estimate of the projected distance between the accreted GC and its host DG. Such a system would fit within the sample of \citet{VanDenBergh2006} who provide a list of 101 GCs with a mean projected distance of $1.62$~kpc to their host DGs.}

We calculate the positions of the two components in integral of motion space since a separation between the two in that space indicates a different infall time between possible progenitors. The procedure and results are displayed in Appendix~\ref{sec:IoMspace} where we observe no separation of this kind with the data we have so far. Although {a GC} scenario is likely, it is still difficult to rule out other potential origins of this stream such as a DG or a NSC accretion scenario. The dynamical perturbation from Sagittarius is also important to include in its history of formation. To have better certainty toward the origin of the Jhelum stream, we would need more member stars with available metallicity and radial velocity measurements.




\section{Conclusions}
\label{sec:conclusion}

In this work, we study the properties of the stellar stream Jhelum, a stream of the Milky Way galaxy that is known for its complex morphology. Our work is based on the findings of \citet{BonacaEtal2019} in which a narrow and a broad component are distinguished as substructures of the stream, and the work of \citet{ShippEtal2019} for which two signals were found in its proper motion space. For this work, we used a recently introduced machine learning methodology, LAAT, to mine these two components and attempt to link their newly estimated properties to the possible merger scenarios that formed the stream. The analysis and results reached in this work can be summarized as follows: 

\begin{itemize}
    \item We used LAAT to highlight the density contrast between the stars more likely to belong to the stream, and stars that are part of the surrounding field. LAAT was applied on four dimensions consisting of two spatial and two proper motion dimensions, and the results were then used to refine the selection of stars in the CMD. 
    
    \item The produced density contrast enhancement revealed two distinct overdensities in proper motion space that we link to the two spatial components of Jhelum: the narrow and broad component. 
    
    \item The separation between these two components was also confirmed a posteriori in radial velocity using measurements from the Southern Stellar Stream Spectroscopic Survey \citep[$S^5$]{LiEtal2019}. We also found that the narrow component has a narrow trend in radial velocity, while the trend for the broad component is more diffuse.
    
    \item With this new information, we calculated properties of the two separated components. Specifically, we used an  MCMC procedure similar to the one used in 
    \citet{WoudenbergEtal2023} to sample from posteriors over the orbits 
    of the two components. This was {done while fitting for the width and velocity dispersion and followed by the calculation of the metallicity dispersion} of either component. For the narrow component, we obtained a velocity dispersion of $4.84^{+1.23}_{-0.79}$~km/s, metallicity dispersion of $0.15^{+0.18}_{-0.10}$, and a width of $28.13^{+8.90}_{-6.64}$~pc. For the broad component, we obtained a velocity dispersion of $19.49^{+2.19}_{-1.84}$~km/s, a metallicity dispersion of $0.34^{+0.13}_{-0.09}$, and a width of $84.09^{+11.26}_{-7.17}$~pc.
    
    \item The small velocity and metallicity dispersion as well as a small width indicate that the narrow component is more likely the result of GC accretion, though it is difficult to fully rule out a DG or a nuclear star cluster progenitor. On the other hand, the comparatively larger dispersions and width of the broad component suggest a DG progenitor for this part of the whole stream. We therefore argue for a likely scenario where Jhelum is the result of the accretion of a GC that belonged to a DG which merged with the Milky Way, although more data are needed to substantiate this claim.
\end{itemize}

It is possible to extend this study by performing deeper medium resolution and/or high resolution follow-up observations of the selection of stars identified in our work. It would be very helpful if these observations would provide high quality metallicity and radial velocity measurements for a larger number of target member stars, as well as light-element abundances (e.g., Na, Mg, and Al), ubiquitous to GCs. 
We leave this attempt therefore for future prospects.

\begin{acknowledgements}
This work is supported by the DSSC Doctoral Training Program of the University of Groningen. We thank the editor for the helpful suggestions that have improved the paper. We thank A. Helmi and T. S. Li for the helpful discussions. We also thank the $S^5$ collaboration for the updated metallicity measurements of their selection of Jhelum members.
EB acknowledges support from a Spinoza prize from the Netherlands
Organisation for Scientific Research (NWO). 
ES acknowledges funding through VIDI grant "Pushing Galactic Archaeology to its limits" (with project number VI.Vidi.193.093) which is funded by the Dutch Research Council (NWO).
We have made use
of data from the European Space Agency (ESA) mission Gaia 
(https://www.cosmos.esa.int/gaia), processed by the Gaia Data Processing and
Analysis Consortium (DPAC,
https://www.cosmos.esa.int/web/gaia/dpac/consortium).  Funding for the DPAC
has been provided by national institutions, in particular the institutions
participating in the Gaia Multilateral Agreement.

\end{acknowledgements}

\bibliographystyle{aa}
\bibliography{references}

\begin{appendix} 

\section{Location in action space}
\label{sec:IoMspace}

\begin{figure}[ht!]
\centering
\includegraphics[width=\linewidth]
{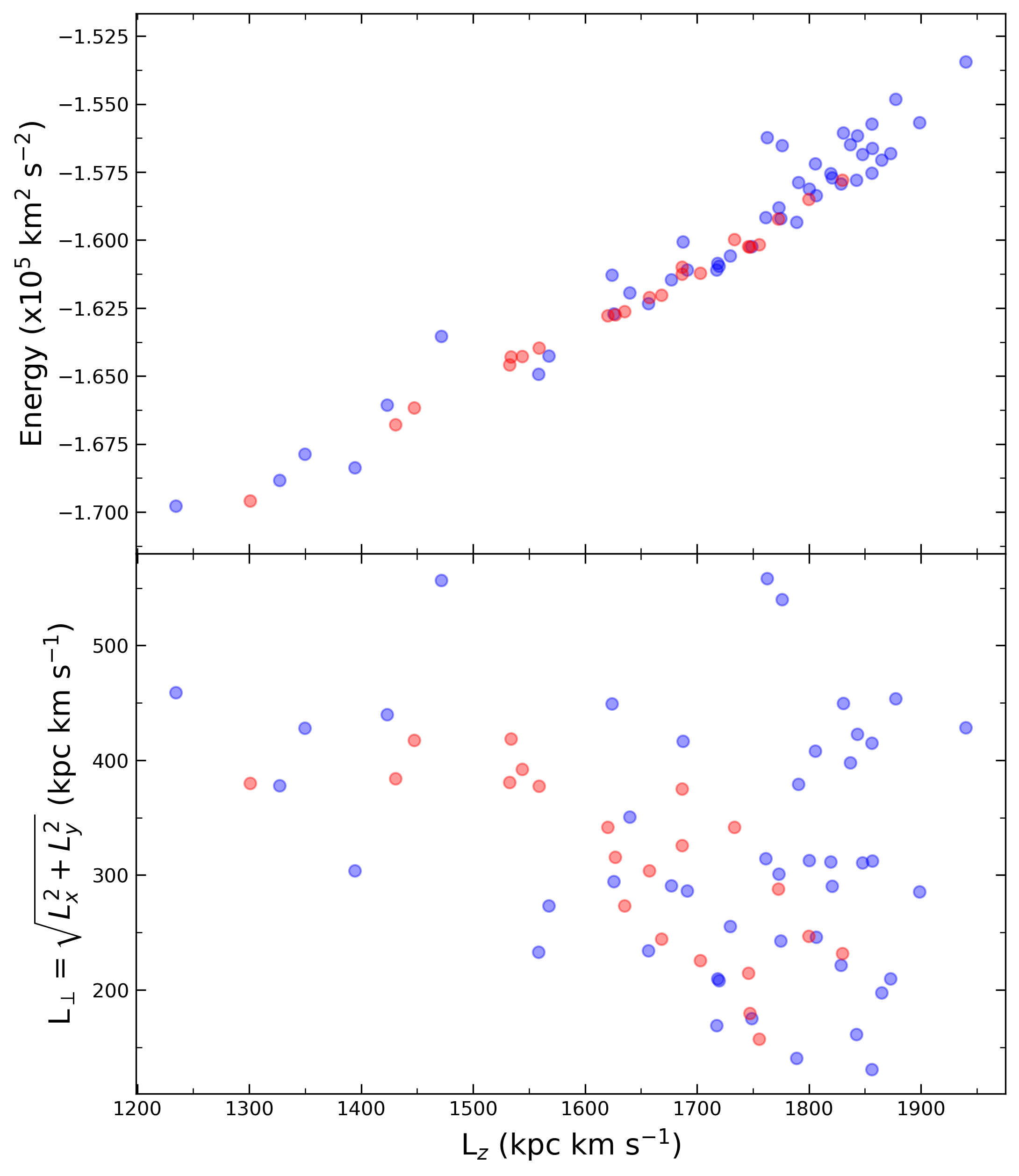}
\caption{Location of the two components in Integral of Motion (IoM) space. In the top panel, we plotted the total energy versus the $z$ component of the angular momentum ($L_z$) for stars belonging to either component of the stream. The lower panel similarly shows the perpendicular component to the total angular momentum $L_{\perp}$ plotted against $L_z$. Though the narrow component seems less scattered in energy versus $L_z$, we see no significant separation between the narrow and broad components in integral of motion space.}
\label{fig:DiscussionFig}
\end{figure}

We test whether there is a difference in integral of motion (IoM) space between the narrow and broad components of Jhelum. We therefore calculate the total energy and angular momentum components of the stars belonging to either component. For this calculation, we use the dynamical parameters we have so far, namely the right ascension and declination as well as the proper motion along either of these coordinates from \emph{Gaia} DR3, and radial velocity measurements from the $S^5$ survey. As for the distances, we make use of the reduced proper motion (RPM) catalog \citep{ViswanathanEtal2023}, where photometric distance measurements of three retrograde streams namely GD-1, Jhelum, and Sagittarius are provided. With this six-dimensional information, we calculate the angular momentum vector and total energy of our sample of stars, assuming the same potential defined in \sref{sec:orbit}. 
We show our results in \fref{fig:DiscussionFig} where we plot the total energy versus $z$-component of the angular momentum ($L_z$) in the top panel, and the perpendicular component versus $L_z$ in the bottom panel. We observe no significant separation between the distribution of stars of the narrow and broad component in the constructed IoM space. This indicates that if the two components of the Jhelum stream have been produced by the accretion two progenitors, they are likely to have been accreted together into the Milky Way.

\section{Completeness and contamination}
\label{sec:completeness}
{In this section we provide a discussion along the lines of completeness and contamination in our selection. In doing so, it is important to keep in mind that providing this estimate would require making assumptions on the starting number of stars belonging to either component of Jhelum and the amount and distribution of contaminating nonmember stars in the field. We construct a mock dataset composed of a stream with two components and of a distribution of surrounding stars. In this way, we would know exactly how many stars belong to either component, how many of them were detected by applying LAAT and enforcing a threshold cutoff, and how many contaminating stars remain at the end of the selection. This approach gives us an idea of the completeness and contamination levels of our sample though in a more simplified setting.}

{To construct the mock catalog, we create three Gaussian distributions, each representing the narrow component, broad component, and the field stars surrounding the stream. Each Gaussian distribution is four dimensional where the dimensions correspond to the two spatial components $(\phi_1, \phi_2)$ and the two proper motion components  $(\mu_{\phi_1}, \mu_{\phi_2})$. The means and covariance matrices of these distributions are set so as to correspond to the positions of the components of the stream in position and proper motion space, as well as to model the distribution of stars along the stream with widths comparable to that of Jhelum. Similarly the distribution corresponding to the field stars is positioned in a way to closely resemble the nonmember stars surrounding the stream. The three distributions would have the following properties for the means and covariance matrices:}

\begin{figure*}[ht]
\centering
\includegraphics[width=\textwidth]{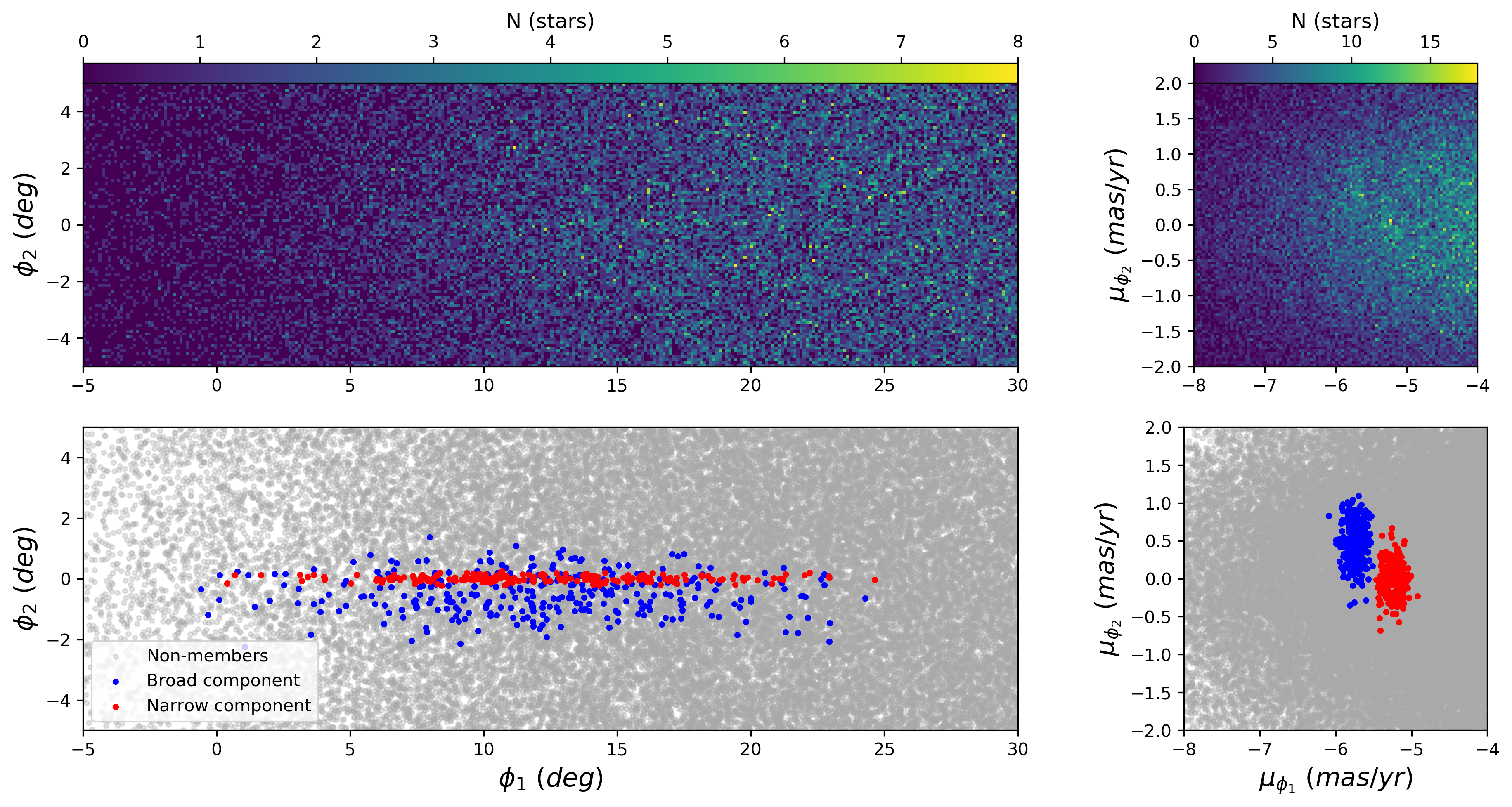}
\caption{{Resulting mock dataset of the Jhelum stream along with a distribution of surrounding field stars. The left panels correspond to the distribution of the sampled stars in position space and the right panels show the location of the same sample in proper motion space. Upper panels: the colors correspond to local density where regions with lighter colors correspond to dense regions. Lower Panels: The generated data points are colored according to their known label with red for the narrow component, blue for broad component and gray for generated nonmembers.}}
\label{fig:SynthData}
\end{figure*}

\begin{figure*}[ht!]
\centering
\includegraphics[width=\textwidth]{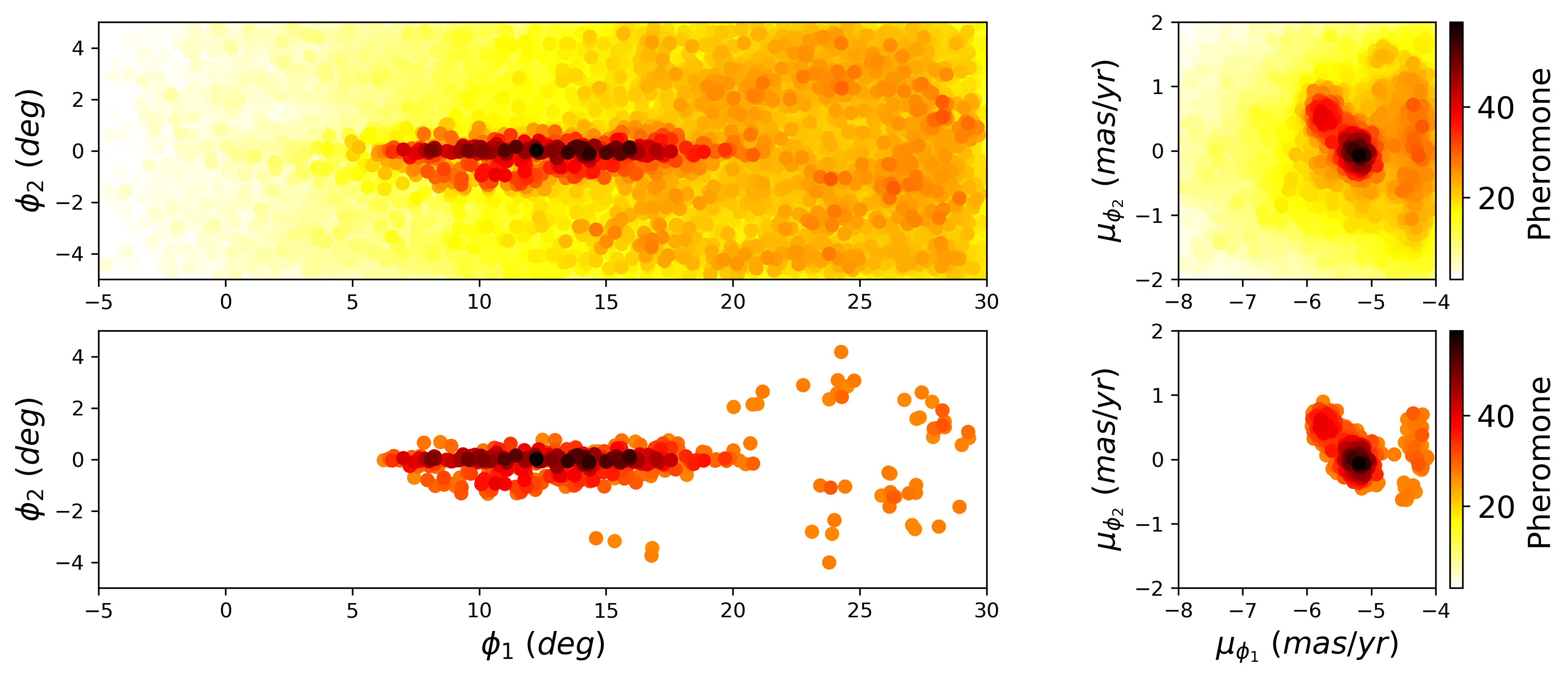}
\caption{{Similar to \fref{fig:LAATRuns} in the paper, we plotted the result of 
LAAT on the four-dimensional space (upper panels), and the result considering a 
threshold of $45\%$ of the maximum value of pheromone during that run (lower panel). 
Since we know the label of each star, evaluating this result gives us an idea of the level of completeness and contamination in our sample.}}
\label{fig:SynthLAAT}
\end{figure*}

{The narrow component $\mathcal{M}^n = (\phi^n_1, \phi^n_2, \mu^n_{\phi_1}, \mu^n_{\phi_2})$, and \\ $\Sigma^n = \mathrm{diag}(\sigma^n_{\phi_1}, \sigma^n_{\phi_2}, \sigma^n_{\mu_{\phi_1}}, \sigma^n_{\mu_{\phi_2}})$, such that $\sigma^n_{\phi_1} > \sigma^n_{\phi_2}$.}

{The broad component $\mathcal{M}^b = (\phi^b_1, \phi^b_2, \mu^b_{\phi_1}, \mu^b_{\phi_2})$, and\\ $\Sigma^b = \mathrm{diag}(\sigma^b_{\phi_1}, \sigma^b_{\phi_2}, \sigma^b_{\mu_{\phi_1}}, \sigma^b_{\mu_{\phi_2}})$, such that $\sigma^b_{\phi_1} > \sigma^b_{\phi_2} > \sigma^n_{\phi_2}$.}

{The field stars $\mathcal{M}^f = (\phi^f_1, \phi^f_2, \mu^f_{\phi_1}, \mu^f_{\phi_2})$, and\\ $\Sigma^f = \sigma^f \times \mathbb{I}$, such that $\sigma^f \gg 1$.} 


{Here $\mathcal{M}$ is the vector of means along each dimension and $\Sigma$ is the covariance matrix given the variances $\sigma^f$ along the diagonal. The matrix $\mathbb{I}$ is the identity matrix, and the superscripts $n$, $b$ and $f$ correspond to the narrow component, broad component and field stars respectively. We now populate this four-dimensional space by sampling stars from these three distributions with proportions similar to those in our sample. In other words, we sample $200$ stars from the narrow component, $300$ stars from the broad component, and $35000$ stars as contaminants. The resulting synthetic dataset can be seen in Figure~\ref{fig:SynthData} where a 2D density plot is provided in the upper row and the true label of each star is shown in the lower row.} 

{We then apply LAAT on this dataset using the same parameters mentioned for Run 1 in \tref{table:1} of the paper. Since we do not have color and magnitude information for this synthetic dataset in order to perform an iterative selection of the stars, we resort instead to applying LAAT once and introducing a harsh threshold on the resulting pheromone distribution. This allows us to place a lower bound on the completeness of the stars since the careful approach to extract the stream introduced in Section~\ref{sec:method} of the paper cannot be followed here, and gives us an idea of the amount of contaminating stars remaining. The results are shown in Figure~\ref{fig:SynthLAAT} of this report where we show the distribution of pheromone after applying LAAT (upper panels) and after applying a threshold of $45\%$ on the maximum amount of pheromone accumulated in that run (lower panels). The threshold is chosen in the same way as explained in Section~\ref{sec:discussion}. It is clear that some contaminant stars remain around the stream which correspond to the stars on the far right side of the proper motion space. These stars would have been further filtered out by the procedure followed in the paper after clustering the two overdensities in proper motion space and applying a cut on the log likelihood to belong to the mixture model (see Section~\ref{sec:results}). Since we know the label of each star in the mock dataset, we can see how many stream stars have been detected and how many contaminating stars remain. Given our results in Figure~\ref{fig:SynthLAAT}, we find that $0.5\%$ of the field stars remain, $80\%$ of the narrow component has been detected, and $34\%$ of the broad component is retrieved.} 

{Though this is a simplified dataset representing the Jhelum stream and its surrounding stars, this experiment gives us an idea of the level of completeness and contamination in our sample. The following claims can be made: the level of contamination in our sample is kept minimal with a contamination percentage of less than $1\%$ of the total stars in the sample. In order to achieve such a level of purity of the selection, member stars of Jhelum are filtered out in the process. This is also primarily due to the global nature of the pheromone threshold where one value is chosen to be applied on all stars of the dataset. In return, diffuse or faint structures run the risk of getting filtered out if more aligned or dense structures are present in the dataset. This effect is clearly seen in this experiment, where we detect $80\%$ of the narrow component but only $34\%$ of the broad component even though the latter is composed of more stars by construction. Since the broad component is more diffuse, our sample is likely missing  more than $50\%$ of the stars belonging to this component. This fact can also be evidenced in the left panel of Figure~4 in the paper as it seems that the region below the narrow component with $\phi_1 > 15^{\circ}$ is missing many of the stars belonging to the broad component. As for the narrow counterpart, its thin and dense nature make it easier to detect with the LAAT algorithm, and so we expect that the sample of stars composing this component is more complete to a level comparable with $80\%$.}

{The fact that we might be missing many of the members of the broad component could mean that we are underestimating some of its properties, for example its width. However, this does not change the conclusions of this paper: the velocity and metallicity dispersion for the broad component clearly point toward a DG origin for this component of the stream. This argument is not applicable to the narrow component however, since we rely on the small dispersions in velocity and metallicity to argue for a GC accretion scenario for this component. That is why it is reassuring that we detect a large portion of the narrow component and that the contamination in our sample is minimal.}

\section{Stream members}
\label{sec:appendixC}

\begin{table*}[ht!]
\small
\caption{List of possible members of the narrow component of Jhelum.}             
\label{table:narrow}      
\centering          
\renewcommand{\arraystretch}{1.25} 
\begin{tabular}{@{}l@{ }c@{ }c@{ }c@{ }c@{ }c@{ }c@{ }c@{ }c@{ }c@{ }c@{ }c@{ }c@{ }c@{}
}     
\hline 
\multicolumn{1}{p{3cm}}{\centering Gaia ID \\ (Gaia DR3)}
&\multicolumn{1}{p{0.9cm}}{\centering RA \\(ICRS)}
&\multicolumn{1}{p{0.9cm}}{\centering Dec \\(ICRS)}
&\multicolumn{1}{p{0.8cm}}{\centering $\mu_{\alpha}$ (mas/ yr)}
&\multicolumn{1}{p{0.8cm}}{\centering $\mu_{\delta}$ (mas/ yr)}
&\multicolumn{1}{p{1.4cm}}{\centering $(BP-RP)_0$ (mag)}
&\multicolumn{1}{p{0.75cm}}{\centering $G_0$ (mag)}
&\multicolumn{1}{p{0.75cm}}{\centering $\varpi$ \\($\prime\prime$)}
&\multicolumn{1}{p{0.75cm}}{\centering $\varpi / \delta_{\varpi}$}
&\multicolumn{1}{c
}{\centering $v_{rad}$ (km/s) }
&\multicolumn{1}{c
}{\centering [Fe/H]}
&\multicolumn{1}{p{1cm}}{\centering $\sigma_{meas}^{[Fe/H]}$}\\
\hline 
 6511703172776180736 & 335.53 & -50.36 & 5.76 & -5.2 & 0.96 & 16.38 & 0.05 & 1.01 & -29.08$^{+1.54}_{-0.77}$ & -1.62 & 0.2 \\
 6511937265674081280 & 334.46 & -50.12 & 5.81 & -5.29 & 0.89 & 17.56 & 0.1 & 1.14 & -21.56$^{+3.72}_{-1.86}$ & -2.14 & 0.35 \\
 6511949016704646144 & 334.08 & -50.01 & 5.65 & -5.33 & 1.04 & 15.47 & 0.08 & 2.39 & -23.7$^{+0.53}_{-0.26}$ & -2.01 & 0.13 \\
 6511950425453887488 & 334.32 & -50.01 & 5.72 & -5.32 & 0.89 & 17.6 & 0.02 & 0.24 & -28.09$^{+1.6}_{-0.81}$ & -1.86 & 0.2 \\
 6512899235269728256 & 344.34 & -51.54 & 6.23 & -4.71 & 0.91 & 17.36 & 0.14 & 1.69 & 5.22$^{+1.1}_{-0.56}$ & -- & -- \\
 6513160536783138816 & 343.29 & -51.1 & 6.06 & -4.53 & 0.63 & 19.01 & 0.27 & 1.31 & 17.29$^{+5.0}_{-2.53}$ & -- & -- \\
 6501420089060434944 & 350.84 & -51.94 & 6.46 & -4.24 & 0.59 & 19.08 & 0.03 & 0.15 & 27.86$^{+7.67}_{-3.47}$ & -- & -- \\
 6501478985948721152 & 350.08 & -51.83 & 6.6 & -4.12 & 0.61 & 18.91 & -0.1 & -0.53 & 23.11$^{+5.85}_{-3.04}$ & -- & -- \\
 6501570825232486400 & 352.47 & -51.97 & 6.43 & -4.12 & 0.69 & 18.67 & -0.08 & -0.53 & 26.82$^{+11.29}_{-5.55}$ & -- & -- \\
 6521621106959698944 & 356.67 & -52.36 & 6.81 & -3.86 & 0.61 & 18.87 & -0.01 & -0.05 & 41.59$^{+5.35}_{-2.73}$ & -- & -- \\
 6521797269338415104 & 359.28 & -52.49 & 6.82 & -3.36 & 0.65 & 18.7 & -0.07 & -0.45 & 46.32$^{+7.18}_{-3.64}$ & -- & -- \\
 6522395201802753024 & 355.87 & -51.97 & 6.68 & -3.72 & 0.92 & 17.19 & 0.02 & 0.28 & 45.88$^{+1.1}_{-0.54}$ & -- & -- \\
 4972241827072899072 & 3.05 & -52.38 & 7.02 & -3.46 & 0.6 & 18.97 & -0.15 & -0.87 & 56.5$^{+9.42}_{-4.78}$ & -- & -- \\
 4972389402151428096 & 1.45 & -52.27 & 7.05 & -3.23 & 0.8 & 18.43 & 0.13 & 1.02 & 53.92$^{+3.02}_{-1.49}$ & -- & -- \\
 4972432622404785152 & 2.73 & -52.34 & 6.81 & -3.03 & 0.61 & 18.84 & 0.01 & 0.09 & 51.36$^{+11.99}_{-5.77}$ & -- & -- \\
 4972434619566888960 & 3.05 & -52.2 & 6.96 & -3.2 & 0.88 & 17.94 & -0.01 & -0.11 & 58.46$^{+2.88}_{-1.43}$ & -- & -- \\
 4973110918000434176 & 0.7 & -52.49 & 6.93 & -3.51 & 0.85 & 18.23 & -0.05 & -0.43 & 43.11$^{+4.65}_{-2.34}$ & -- & -- \\
 4973123184427063296 & 0.06 & -52.42 & 7.04 & -3.66 & 0.73 & 18.62 & -0.17 & -1.21 & 47.64$^{+5.01}_{-2.49}$ & -- & -- \\
 4973132259695258624 & 0.53 & -52.19 & 6.97 & -3.41 & 0.9 & 17.41 & -0.13 & -1.77 & 50.36$^{+2.3}_{-1.15}$ & -1.76 & 0.15 \\
 4973136172408386560 & 0.9 & -52.36 & 6.94 & -3.54 & 0.73 & 18.55 & -0.03 & -0.2 & 49.19$^{+4.33}_{-2.24}$ & -- & -- \\
 4973136481646042112 & 0.94 & -52.32 & 6.82 & -3.06 & 0.63 & 18.92 & -0.04 & -0.25 & 49.38$^{+8.99}_{-4.38}$ & -- & -- \\
 4973148507554608128 & 1.41 & -52.07 & 7.05 & -2.92 & 0.65 & 18.75 & 0.02 & 0.12 & 56.54$^{+6.5}_{-3.1}$ & -- & -- \\
\hline   
\end{tabular}
\tablefoot{The columns from left to right correspond to the following: the Gaia DR3 ID, current positions on the sky, proper motions, colors, magnitudes, parallax, relative error on parallax, line-of-sight velocities, $S^5$ metallicity, and error on the metallicity.}
\end{table*}


\begin{table*}[ht!]
\small
\caption{List of possible members of the broad component of Jhelum.}             
\label{table:broad}      
\centering  
 \renewcommand{\arraystretch}{1.15} 
\begin{tabular}{@{}l@{ }c@{ }c@{ }c@{ }c@{ }c@{ }c@{ }c@{ }c@{ }c@{ }c@{ }l@{ }c@{ }c@{}}     
\hline 

\multicolumn{1}{p{3cm}}{\centering Gaia ID \\(Gaia DR3)}
&\multicolumn{1}{p{0.9cm}}{\centering RA \\(ICRS)}
&\multicolumn{1}{p{0.9cm}}{\centering Dec \\(ICRS)}
&\multicolumn{1}{p{0.8cm}}{\centering $\mu_{\alpha}$ (mas/ yr)}
&\multicolumn{1}{p{0.8cm}}{\centering $\mu_{\delta}$ (mas/ yr)}
&\multicolumn{1}{p{1.4cm}}{\centering $(BP-RP)_0$ (mag)}
&\multicolumn{1}{p{0.75cm}}{\centering $G_0$ (mag)}
&\multicolumn{1}{p{0.75cm}}{\centering $\varpi$ \\($\prime\prime$)}
&\multicolumn{1}{p{0.75cm}}{\centering $\varpi / \delta_{\varpi}$}
&\multicolumn{1}{c
}{\centering $v_{rad}$ (km/s) }
&\multicolumn{1}{c
}{\centering [Fe/H]}
&\multicolumn{1}{p{1cm}}{\centering $\sigma_{meas}^{[Fe/H]}$}\\
\hline 
 6563748345221978112 & 323.53 & -46.7 & 5.24 & -6.45 & 0.54 & 19.1 & 0.42 & 1.53 & -63.65$^{+9.3}_{-4.5}$ & -- & -- \\
 6513551864140520448 & 341.73 & -51.33 & 6.92 & -5.64 & 0.96 & 16.77 & 0.09 & 1.53 & -60.09$^{+8.16}_{-4.05}$ & -- & -- \\
 6513586464397103104 & 341.88 & -51.01 & 6.77 & -5.29 & 0.91 & 16.99 & -0.03 & -0.47 & -28.71$^{+0.93}_{-0.46}$ & -1.77 & 0.15 \\
 6513595702869676032 & 341.46 & -50.96 & 6.83 & -5.12 & 0.65 & 18.75 & 0.11 & 0.56 & -43.4$^{+3.59}_{-1.88}$ & -- & -- \\
 6513741559959126016 & 339.46 & -50.84 & 6.58 & -5.64 & 0.8 & 18.43 & 0.15 & 0.97 & -53.12$^{+4.72}_{-2.45}$ & -- & -- \\
 6513756506444306432 & 340.92 & -51.09 & 6.89 & -5.57 & 0.85 & 18.28 & -0.07 & -0.45 & -55.67$^{+6.22}_{-3.02}$ & -- & -- \\
 6513756957417138176 & 340.85 & -51.04 & 6.63 & -5.29 & 0.89 & 17.57 & -0.03 & -0.27 & -26.71$^{+3.52}_{-1.78}$ & -2.36 & 0.53 \\
 6511069785359880192 & 336.54 & -49.96 & 6.48 & -5.97 & 0.75 & 18.56 & 0.07 & 0.45 & -75.17$^{+9.17}_{-4.63}$ & -- & -- \\
 6513917932791132160 & 342.04 & -50.81 & 6.91 & -5.85 & 0.99 & 16.34 & 0.05 & 0.93 & -51.97$^{+0.42}_{-0.22}$ & -1.58 & 0.16 \\
 6514001358235953152 & 342.1 & -50.33 & 6.94 & -5.69 & 0.98 & 15.97 & 0.06 & 1.42 & -29.63$^{+0.38}_{-0.2}$ & -2.19 & 0.16 \\
 6514110003728640000 & 342.19 & -49.92 & 6.93 & -5.2 & 0.78 & 18.35 & -0.16 & -1.04 & -20.06$^{+6.95}_{-3.5}$ & -- & -- \\
 6516954092417230848 & 338.56 & -49.4 & 6.64 & -5.8 & 0.83 & 18.24 & 0.15 & 1.14 & -43.74$^{+3.2}_{-1.58}$ & -- & -- \\
 6516999382848036864 & 338.0 & -50.22 & 6.22 & -5.64 & 0.79 & 18.49 & 0.32 & 1.98 & -56.12$^{+4.51}_{-2.25}$ & -- & -- \\
 6517085282194434048 & 336.91 & -49.63 & 6.64 & -5.64 & 0.79 & 17.97 & 0.18 & 1.41 & -33.68$^{+2.14}_{-1.07}$ & -- & -- \\
 6514759540222683136 & 344.28 & -50.09 & 7.13 & -5.28 & 0.85 & 17.29 & 0.19 & 2.26 & 16.25$^{+1.82}_{-0.9}$ & -- & -- \\
 6512875149093073920 & 344.85 & -51.91 & 7.26 & -5.13 & 0.92 & 17.64 & 0.44 & 4.74 & 5.94$^{+1.32}_{-0.64}$ & -- & -- \\
 6512895932437355520 & 344.38 & -51.68 & 6.93 & -5.63 & 0.76 & 18.46 & 0.39 & 2.6 & -49.0$^{+3.99}_{-1.98}$ & -- & -- \\
 6512934003027946496 & 345.32 & -51.51 & 6.86 & -5.29 & 0.77 & 18.44 & 0.08 & 0.55 & -34.32$^{+3.8}_{-1.85}$ & -- & -- \\
 6512998397473564672 & 344.72 & -51.33 & 7.34 & -5.84 & 0.88 & 17.38 & 0.06 & 0.69 & -45.58$^{+1.65}_{-0.84}$ & -2.04 & 0.17 \\
 6513114219855440896 & 344.01 & -51.05 & 7.13 & -5.91 & 0.68 & 19.38 & 0.01 & 0.04 & -35.06$^{+13.58}_{-6.88}$ & -- & -- \\
 6513134423381780480 & 343.16 & -51.26 & 7.27 & -5.67 & 0.61 & 19.1 & -0.1 & -0.43 & -37.76$^{+6.02}_{-3.06}$ & -- & -- \\
 6513228023604300800 & 344.56 & -50.72 & 7.1 & -4.99 & 0.66 & 19.0 & -0.2 & -0.82 & -6.72$^{+7.73}_{-3.93}$ & -- & -- \\
 6513240560613556224 & 344.34 & -50.43 & 6.97 & -4.94 & 0.7 & 18.44 & 0.42 & 2.54 & -36.51$^{+7.45}_{-3.84}$ & -- & -- \\
 6513269422793502720 & 343.8 & -50.66 & 6.77 & -5.23 & 0.61 & 19.45 & -0.12 & -0.35 & -69.91$^{+48.97}_{-7.77}$ & -- & -- \\
 6513263615997686784 & 343.71 & -50.69 & 7.16 & -5.33 & 0.59 & 19.01 & -0.02 & -0.07 & -33.89$^{+12.25}_{-6.02}$ & -- & -- \\
 6513301759602918400 & 343.54 & -50.26 & 6.98 & -5.13 & 0.85 & 16.99 & 0.21 & 3.07 & -6.16$^{+1.12}_{-0.54}$ & -- & -- \\
 6516771371624716288 & 339.12 & -50.41 & 6.43 & -5.77 & 0.96 & 16.18 & 0.02 & 0.38 & -57.6$^{+0.53}_{-0.26}$ & -1.97 & 0.12 \\
 6516823113094551552 & 338.85 & -50.25 & 6.71 & -5.9 & 0.6 & 18.75 & 0.32 & 1.92 & -61.3$^{+4.36}_{-2.2}$ & -- & -- \\
 6516837922141932544 & 338.21 & -50.08 & 6.28 & -5.87 & 0.56 & 18.91 & -0.16 & -0.72 & -43.23$^{+7.46}_{-3.76}$ & -- & -- \\
 6502177274615226368 & 347.8 & -51.69 & 7.23 & -5.1 & 0.59 & 18.97 & -0.09 & -0.46 & -33.16$^{+6.63}_{-3.21}$ & -- & -- \\
 6502215761817932800 & 349.38 & -51.83 & 7.65 & -5.3 & 0.65 & 18.53 & 0.02 & 0.12 & -27.58$^{+6.8}_{-3.51}$ & -- & -- \\
 6502315538202150912 & 348.05 & -51.04 & 7.27 & -4.7 & 0.7 & 19.34 & -0.14 & -0.45 & -19.26$^{+15.66}_{-6.68}$ & -- & -- \\
 6502377488810816512 & 346.96 & -51.58 & 7.51 & -5.1 & 0.65 & 19.21 & 0.21 & 0.97 & -37.98$^{+8.95}_{-4.6}$ & -- & -- \\
 6502384841794895872 & 347.13 & -51.48 & 7.47 & -5.42 & 0.59 & 19.17 & -0.04 & -0.19 & -23.36$^{+6.18}_{-2.94}$ & -- & -- \\
 6500871707636389888 & 346.96 & -51.87 & 7.21 & -5.26 & 0.54 & 19.16 & -0.39 & -1.67 & -15.64$^{+9.35}_{-4.75}$ & -- & -- \\
 6500910504077886464 & 346.09 & -51.69 & 7.19 & -5.08 & 0.94 & 17.06 & 0.03 & 0.44 & -30.97$^{+2.08}_{-1.0}$ & -1.83 & 0.28 \\
 6500936613181406208 & 346.1 & -51.53 & 6.79 & -4.91 & 0.66 & 18.66 & 0.43 & 2.86 & -30.92$^{+11.73}_{-5.93}$ & -- & -- \\
 6501415106898347008 & 351.14 & -52.04 & 7.64 & -5.06 & 0.6 & 18.91 & -0.2 & -1.13 & -48.67$^{+34.96}_{-18.78}$ & -- & -- \\
 6501458404465461248 & 349.64 & -52.04 & 7.46 & -5.1 & 1.05 & 15.56 & 0.04 & 1.51 & -14.86$^{+0.44}_{-0.21}$ & -1.58 & 0.12 \\
 6501475653054097408 & 350.2 & -51.88 & 7.61 & -5.29 & 0.95 & 17.18 & 0.44 & 6.75 & -75.07$^{+1.32}_{-0.67}$ & -- & -- \\
 6501507294076007424 & 349.71 & -51.38 & 7.03 & -4.65 & 0.74 & 18.36 & 0.13 & 0.83 & -3.18$^{+2.68}_{-1.35}$ & -- & -- \\
 6501641232632969216 & 351.65 & -51.61 & 7.34 & -4.69 & 0.68 & 18.95 & 0.11 & 0.62 & -7.27$^{+9.69}_{-4.93}$ & -- & -- \\
 6501725237896628224 & 351.95 & -51.45 & 7.64 & -4.72 & 0.87 & 17.7 & -0.06 & -0.62 & -18.46$^{+4.75}_{-2.39}$ & -1.75 & 0.35 \\
 6501799149989898240 & 350.87 & -51.68 & 7.69 & -5.1 & 0.96 & 16.83 & 0.13 & 2.37 & -10.76$^{+0.83}_{-0.41}$ & -1.1 & 0.13 \\
 6501804853706465280 & 350.88 & -51.67 & 7.71 & -5.1 & 0.85 & 18.04 & 0.09 & 0.77 & -18.49$^{+2.33}_{-1.13}$ & -- & -- \\
 6501846390334223360 & 350.82 & -51.24 & 7.39 & -4.45 & 0.84 & 17.13 & 0.32 & 4.76 & -56.91$^{+1.07}_{-0.55}$ & -- & -- \\
 6498579333267972096 & 353.79 & -52.27 & 7.5 & -4.56 & 0.89 & 17.06 & -0.02 & -0.39 & -6.99$^{+1.23}_{-0.6}$ & -- & -- \\
 6501947992080634880 & 351.06 & -50.78 & 7.6 & -4.78 & 0.81 & 18.33 & 0.03 & 0.21 & 14.61$^{+8.31}_{-4.02}$ & -- & -- \\
 6560094019545230336 & 333.63 & -49.13 & 6.3 & -6.04 & 0.83 & 17.99 & 0.2 & 1.6 & -83.51$^{+6.87}_{-3.63}$ & -- & -- \\
 6560348242952310784 & 332.23 & -48.54 & 6.11 & -6.13 & 0.82 & 18.12 & -0.06 & -0.51 & -91.9$^{+12.2}_{-5.96}$ & -- & -- \\
 6525718883717574656 & 353.04 & -51.07 & 7.28 & -4.61 & 0.63 & 18.94 & -0.03 & -0.14 & 25.42$^{+15.47}_{-7.13}$ & -- & -- \\
 6522488969528625152 & 356.37 & -51.74 & 7.89 & -4.59 & 0.79 & 18.01 & 0.18 & 1.74 & 11.93$^{+1.97}_{-0.99}$ & -- & -- \\
 6522627782871271424 & 354.35 & -51.71 & 7.74 & -4.4 & 0.7 & 18.57 & 0.19 & 1.08 & 12.71$^{+4.79}_{-2.41}$ & -- & -- \\
 4972463683609076736 & 3.15 & -51.82 & 7.98 & -3.68 & 0.62 & 18.76 & 0.0 & 0.01 & 32.38$^{+5.85}_{-2.85}$ & -- & -- \\
 
\hline    
\end{tabular}
\tablefoot{The columns from left to right correspond to the follwoing: the Gaia DR3 ID, current positions on the sky, proper motions, colors, magnitudes, parallax, relative error on parallax, line-of-sight velocities, $S^5$ metallicity, and error on the metallicity.}
\end{table*}
\end{appendix}
\end{document}